\newcommand{\teff}{T$_{\mathrm{eff}}$}

\newcommand{\logg}{log~$g$}
\newcommand{\feh}{[Fe/H]}
\newcommand{\kms}{km~s$^{-1}$}

\newcommand{\vmicro}{$\xi$}

\newcommand{\numstar}{50} 
\newcommand{\numpair}{25} 
\newcommand{\numel}{23} 

\newcommand{\project}[1]{\textsl{#1}}
\newcommand{\gaia}{\project{Gaia}}

\documentclass[a4paper,fuseAMS,leqn,usenatbib]{mnras}

\usepackage{graphicx}
\usepackage{amssymb}
\usepackage{amsmath}
\usepackage{times}
\usepackage{subfigure}
\usepackage{float}
\usepackage{longtable}
\usepackage[T1]{fontenc}
\usepackage{ae,aecompl}

\title[Chemical Homogeneity of Twin Stars]{Identical or fraternal twins? : The chemical homogeneity of wide binaries from \gaia\ DR2}

 \author[Hawkins et al. 2019]{Keith~Hawkins$^{1}$\thanks{E-mail: keithhawkins@utexas.edu}, Madeline~Lucey$^{1}$, Yuan-Sen Ting$^{2, 3, 4,\dagger}$,  Alexander Ji$^{4,\dagger}$,\newauthor  Dustin~Katzberg$^{1}$ , Megan~Thompson$^{1}$, Kareem El-Badry$^{5}$, Johanna Teske$^{4,\dagger}$, \newauthor Tyler Nelson$^{1}$, Andreia Carrillo$^{1,\ddagger}$\\
$^{1}$Department of Astronomy, The University of Texas at Austin, 2515 Speedway Boulevard, Austin, TX 78712, USA\\
$^{2}$Institute for Advanced Study, Princeton, NJ 08540, USA\\
$^{3}$Department of Astrophysical Sciences, Princeton University, Princeton, NJ 08544, USA\\
$^{4}$Observatories of the Carnegie Institution of Washington, 813 Santa Barbara Street, Pasadena, CA 91101, USA \\
$^{5}$Department of Astronomy and Theoretical Astrophysics Center, University of California Berkeley, Berkeley, CA 94720\\
$^{\dagger}$ NASA Hubble Fellow\\
$^{\ddagger}$ LSSTC Data Science Fellow\\
 }

\date{Accepted XX. Received YY; in original form ZZ}

\pubyear{2019}

\begin{document}
\label{firstpage}
\pagerange{\pageref{firstpage}--\pageref{lastpage}}
\maketitle

\begin{abstract}
One of the high-level goals of Galactic archaeology is chemical tagging of stars across the Milky Way to piece together its assembly history. For this to work, stars born together must be uniquely chemically homogeneous. Wide binary systems are an important laboratory to test this underlying assumption. Here we present the detailed chemical abundance patterns of \numstar\ stars across \numpair\ wide binary systems comprised of main-sequence stars of similar spectral type identified in \gaia~DR2 with the aim of quantifying their level of chemical homogeneity. Using high-resolution spectra obtained with McDonald Observatory, we derive stellar atmospheric parameters and  precise detailed chemical abundances for light/odd-Z (Li, C, Na, Al, Sc, V, Cu), $\alpha$ (Mg, Si, Ca), Fe-peak (Ti, Cr, Mn, Fe, Co, Ni, Zn), and neutron capture (Sr, Y, Zr, Ba, La, Nd, Eu) elements. Results indicate that 80\% (20 pairs) of the systems are homogeneous in \feh\ at levels below 0.02~dex. These systems are also chemically homogeneous in all elemental abundances studied, with offsets and dispersions consistent with measurement uncertainties. We also find that wide binary systems are far more chemically homogeneous than random pairings of field stars of similar spectral type. These results indicate that wide binary systems tend to be chemically homogeneous but in some cases they can differ in their detailed elemental abundances at a level of [X/H] $\sim$0.10~dex, overall implying chemical tagging in broad strokes can work. 

\end{abstract}


\section{Introduction}
\label{sec:Introduction}
Chemical tagging is among one of the more popular and high-level goals of modern Galactic archaeology. The power behind this technique, proposed nearly two decades ago \citep{Freeman2002}, is that it asserts that we can determine the birth place of stars given their chemical composition alone.  If possible, chemical tagging would enable us to both identify dispersed stellar clusters and accreted material. This makes chemical tagging a uniquely powerful tool to reconstruct the formation and evolutionary history of the Galaxy. The possibility of being able to carry out chemical tagging on an industrial level is one of the core motivations for investments in large spectroscopic surveys, which include the GALactic Archaeology with HERMES \citep[GALAH,][]{De_silva2015} survey, the Apache Point Observatory Galactic Evolution Experiment \citep[APOGEE,][]{Majewski2017}, the Large Sky Area Multi-Object Fibre Spectroscopic Telescope survey \citep[LAMOST,][]{Luo2015, Xiang2017} and the Radial VElocity Experiment \citep[RAVE,][]{Kunder2017}. While the prospects of chemical tagging is promising, doing it in practice is challenging \citep[e.g.][]{DeSilva2007, Mitschang2014, Ting2015, Bovy2016, Hogg2016, Kos2016}. This is partly because it is not clear whether the underlying assumptions of the technique are valid across the Milky Way \citep[e.g.][]{Ness2018}

In order for chemical tagging to work, a few assumptions must be satisfied. Namely, it is required that stars that form in pairs, groups, or clusters are (i) chemically homogeneous, and  (ii) unique from other groups. That is to say, while over time, the stars that formed in a given group or cluster may disperse spatially or kinematically, they will continue to belong to the chemically unique group that they were formed in.  In this context, wide binaries are one of the best laboratories for testing the validity of these key assumptions that underpin chemical tagging. 

Wide binaries are thought to be formed in a variety of ways. Those with separations between a hundred and a few thousand AU are thought to form primarily through turbulent core fragmentation \citep[e.g.][]{Offner2010, Lee2017}. Binaries with even wider separations, 0.01 to 1~pc, have been proposed to form through dynamical evolution of unstable triples \citep{Reipurth2012}, the dissolution of star clusters \citep[e.g.][]{Kouwenhoven2010, Moeckel2011}, or pairing of dynamically adjacent cores \citep[e.g.][]{Tokovinin_2017}. In most formation channels for wide binary systems, they are formed at approximately the same time (coeval) and from the same gas (co-natal). These two points make wide binaries not only useful to test the underlying assumptions of chemical tagging but have many additional astrophysical applications. 
\\ \\ 
For example, wide binaries are often used for the calibration of the atmospheric and chemical parameters of stars that are difficult to analyze. M-dwarf stars have low enough temperatures (\teff\ $<$ 4000~K) that their spectra contain many molecular features making them difficult to characterize. However, the metallicity (and chemical composition) of M-dwarfs can be determined if they have a wide binary companion that is easier to analyze \citep[e.g.][]{Mann2013, Lepine2007, Rojas-Ayala2010, Montes2018}, though temperature-dependent settling of metals in stellar atmospheres can complicate this \citep{Dotter2017}. Wide binaries containing white dwarfs have been used to measure the ages of main-sequence companions \citep{Chaname2012, Fouesneau2019} and determine the metallicity of the white dwarf's progenitor, which is useful for constraining the initial-final mass relation \citep[e.g.][]{Zhao2012, Andrews2015}. Beyond these, there are many other applications of wide binary stellar systems discussed in the literature \citep[e.g.][]{Poveda1994, Bahcall1985, Yoo2004, Shaya2011, Garces2011, Chaname2012, Tokovinin2012, Alonso-Floriano2015, Penarrubia2016, ElBadry2018a, ElBadry2019a}. 

Critically, many of the applications of wide binaries rely on them being chemically homogeneous, co-natal, and coeval systems. This is expected based on early results  \citep[e.g.][]{Gizis1997, Gratton2001, Martin2002, Desidera2004, Desidera2006}. However, \cite{Oh2018}, while exploring the detailed chemical abundances of wide binaries using high resolution spectra from \cite{Brewer2016}, found an example of wide binary systems where the metallicity (and other elements) differed as much as 0.20~dex, in a pattern that suggested accretion from rocky planetary material. This, however had been seen in several previous earlier studies on other systems which include: 16 Cygni \citep[e.g.][with variation in metallicity between the two on the order of 0.04 dex]{Laws2001, Ramirez2011b,TucciMaia2014}, XO-2 \citep[e.g.][where the metallicity can vary between the pairs by as much as 0.10 dex]{Biazzo2015, Ramirez2015}, the WASP-94 system \citep[e.g.][who found differences in the metal content of the binary pair at the level of 0.02 dex]{Teske2016}, and the HAT-P-4 system \citep[e.g.][who found 0.10 dex difference in the metallicity between the two component of this binary]{Saffe2017}. More recently, \cite{Ramirez2019}, showed that there is a significant difference ($\Delta$\feh\ $\sim$ 0.17~dex) in the \feh\ and other elemental abundance ratios in the wide binary system HD34407-HD34426. For more discussion on the impact of these systems we refer the reader to the annual review of \cite{Nissen2018} and references therein. Each of these studies found differences between wide binaries ranging in size from 0.01 -- 0.20~dex. These results, along with other recent works \citep[e.g.][]{Simpson2018}, raised the question whether significant chemical variation between the components of wide binaries is common or unusual. 

In the last couple of years, there has been much discussion in the literature centered on the identification and characterization of wide binary systems \citep[e.g.][]{Andrews2017, Oh2017, Oh2018, Oelkers2017, Price-Whelan2017, ElBadry2018a, Simpson2018, Andrews2019}.  High precision parallaxes and proper motions from the second release of the \gaia\ mission \citep[\gaia~DR2,][]{Gaiasummary2018} have recently made it straightforward to construct large samples of  high-confidence wide binaries \citep[e.g.][]{ElBadry2018a}. With these newly discovered systems we are now in a position to begin to determine the level to which wide binaries are chemically identical or fraternal, thereby testing the fundamental assumptions of chemical tagging. 

In this work, we preform a detailed chemical abundance analysis of a sample of wide binaries identified in \gaia~DR2 covering range of separations in order to quantify the co-natal, homogeneous assumption of chemical tagging. In order to do this, in section~\ref{subsec:selection} we discuss the selection of co-moving pairs from \cite{ElBadry2018a}. We observed a subsample of these co-moving pairs and discuss the properties of the spectral data obtained in section~\ref{subsec:spec}. In section~\ref{sec:SP}, we outline the process used to derive the stellar atmospheric parameters and detailed chemical abundance from the observational data. The results of this work in the context of recent literature on the chemical homogeneity of wide binaries is presented in section~\ref{sec:results}. Finally,  in section~\ref{sec:summary}, we summarize our results, showing that co-moving wide binary systems are chemically homogeneous at a level below 0.08~dex across all 24 elements studied. 

\section{Data} \label{sec:data}
\subsection{Selecting Co-Moving Pairs from Gaia DR2}
\label{subsec:selection}
\begin{figure}
	 \includegraphics[width=1\columnwidth]{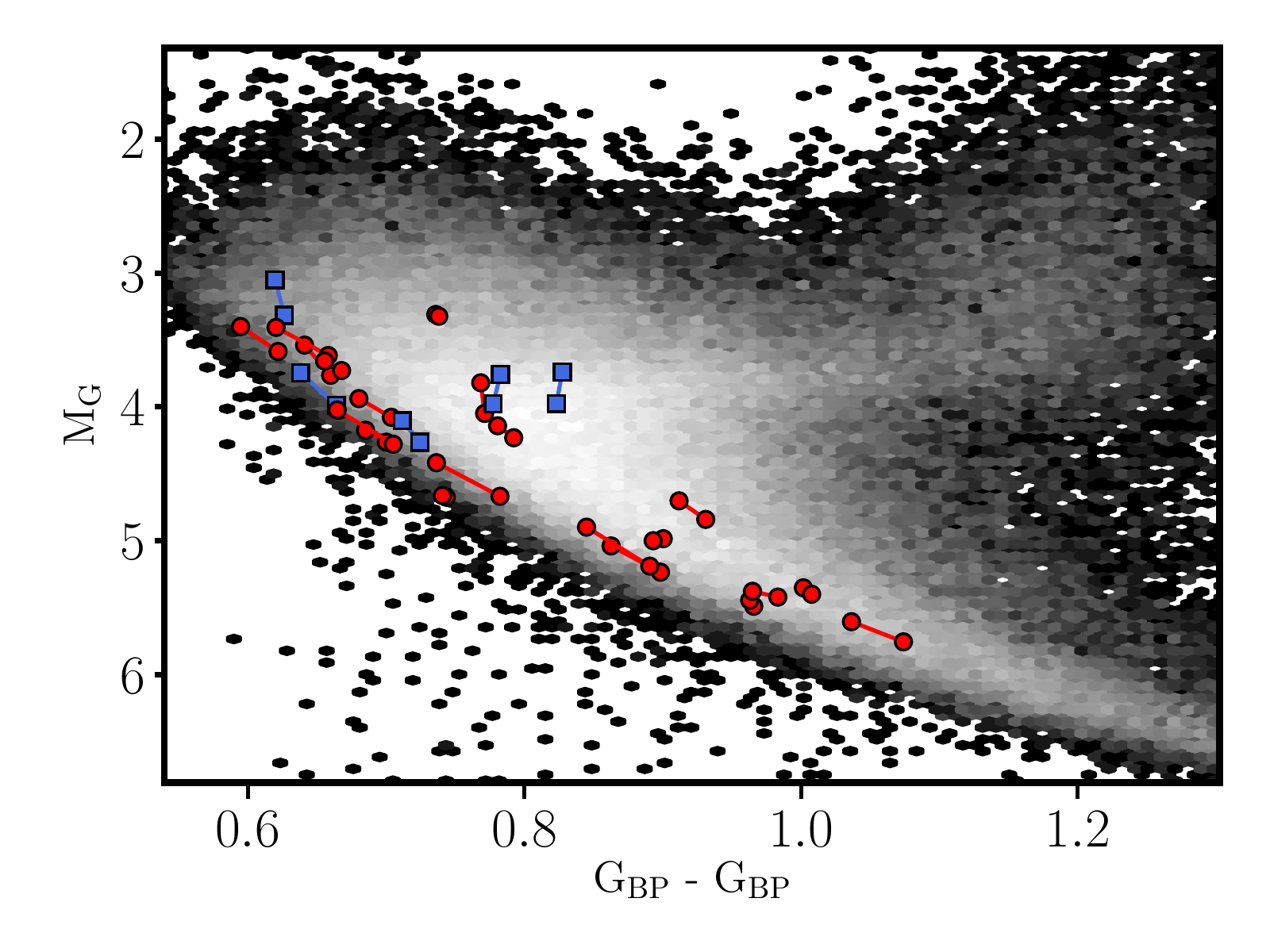}
	\caption{The absolute magnitude in the $G$ band, M$_{G}$, as a function of colour ($BP-RP$) for the observed co-moving pairs which turn out to be chemically homogeneous (red circles where pairs are connected by solid lines) and those which have $\Delta$\feh\ larger than 0.10~dex (blue squares where pairs are connected by solid lines). For reference, the absolute magnitude in the $G$ band as a function of colour ($BP-RP$) of the GALAH survey crossmatch with \gaia~DR2 (with parallax uncertainties better than 10\% and parallaxes larger than 1) is also shown as the grayscale background. } 
	\label{fig:CMD}
\end{figure}

In order to determine the level of chemical homogeneity in co-moving binary stellar systems, we started with the set of main-sequence/main-sequence (MS/MS) co-moving pairs identified in \cite{ElBadry2018a}. We summarize the method of these authors here. In \cite{ElBadry2018a}, the authors identified $\sim50 \times 10^4$~MS/MS wide binaries within 200~pc with projected separations between 50~$< s <$~50000~AU with less than 1\% contamination. After doing an initial quality control cut on the stars within \gaia\ identified with distances less than 200~pc, they do an initial search for companions around each star by (i) rejecting any companions whose parallaxes were inconsistent with that of the primary at the 3-sigma level, and (ii) requiring that the difference in the proper motions of the two stars in the pair be consistent with a bound Keplerian orbit. The authors then removed clusters, moving groups and higher-order multiples outside of pairs of two stars. For a more detailed discussion on the identification of wide binaries, the removal of higher-order multiples, and the expected contamination rate we refer the reader to section~2 of \cite{ElBadry2018a}.

We applied several additional cuts. In order to focus in this work on stars with similar \teff, as a way to reduce potential systematics in the derived parameters and abundances \citep[e.g.][]{Andrews_2019}, we required the difference between the $G$ magnitude of both stars in the pair to be less than 0.30~mag and the difference in the $G_{BP}-G_{RP}$ color to also be less than 0.05~mag. The majority of these pairs are part of the excess of photometric ``twin'' binaries with mass ratios near 1 discussed in \citet{ElBadry2019b}. This led to an initial sample of 2948 stars across 1474 co-moving pairs. Of these stars, we were able to observe \numstar\ stars across \numpair\ co-moving pairs at McDonald Observatory in January 2019 (more details in section~\ref{subsec:spec}). They were selected by prioritizing the bright stars while trying to span a range of projected separations. They were also selected to be far enough apart to minimize light from the companion entering the slit. These stars are typically brighter than G$\sim$ 12~mag. A color magnitude diagram (in M$_{G}$ as a function of $BP-RP$ for the observed co-moving pairs (red and blue circles) can be found in Fig.~\ref{fig:CMD}.

%

\subsection{High Resolution Spectra From McDonald Observatory}
\label{subsec:spec}
\begin{figure*}
	 \includegraphics[width=2\columnwidth]{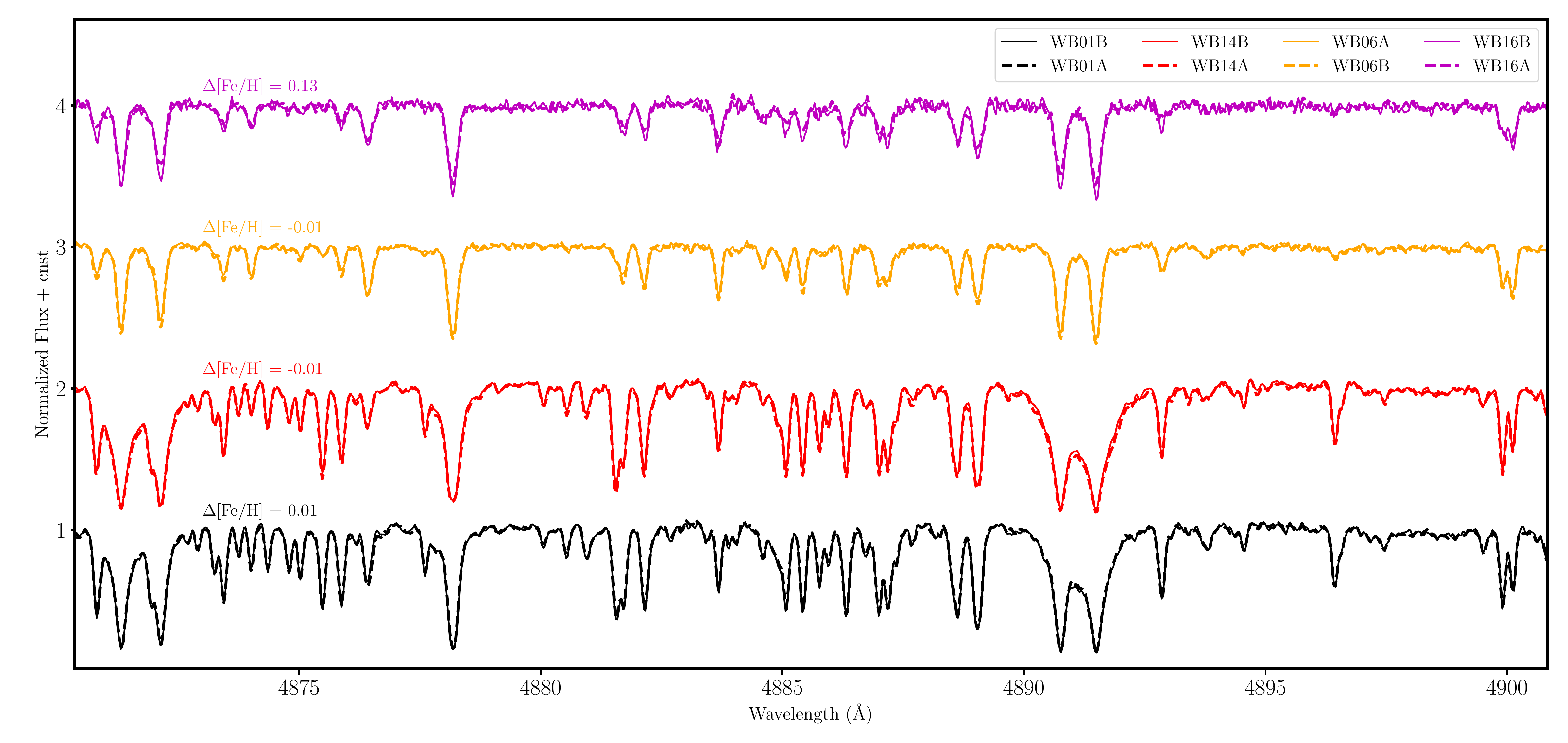}
	\caption{Here we show the observed spectra in the spectra region between 4870--4905~\AA\ of 4 wide binary systems shown as different colours. One component of the binary system is shown as a dotted line, while its companion is shown as a solid line. The difference in \feh\ between these spectra are also shown. The spectra of these representative wide binary systems are remarkably similar, except for WB16. As is shown this is one pair where the metallicity is different between the two stars by 0.13~dex } 
	\label{fig:spectra}
\end{figure*}

In order to quantify the level of chemical homogeneity, we observed \numstar\  stars across \numpair\ co-moving pairs initially identified in \cite{ElBadry2018a} with the Tull Echelle Spectrograph \citep{Tull1995} on the 2.7m Harlan J. Smith Telescope at McDonald Observatory in early 2019. The sample size was selected to be comparable to current studies of co-moving pairs of stars \citep[e.g.][]{Oh2018, Simpson2018, Andrews2019}. These observations enabled us to obtain high-resolution (with a resolving power of R $=\lambda/\Delta\lambda \sim$ 60000) optical spectra. We also obtained standard calibration exposures (i.e. biases, flats, and wavelength comparison, ThAr, lamps). The spectra were reduced in the standard way including subtraction of the bias, dividing by the flat field, optimal spectra extraction and scattered light subtraction.  In order to stitch the various echelle orders together, we did an initial continuum normalization assuming a fifth order spline function. These processes were done using the echelle package with IRAF\footnote{IRAF is distributed by the National Optical Astronomy Observatory, which is operated by the Association of Universities for Research in Astronomy (AURA) under a cooperative agreement with the National Science Foundation.}. Radial Velocity (RVs) for the spectra were determined by cross correlation with a solar spectral template with the iSpec package \citep{2014A&A...569A.111B}. If multiple spectra were observed for the same target, these spectra were co-added (after barycentric correction) in order to obtain the highest SNR possible. For all 25 pairs, the RVs, reported in Table~\ref{tab:obsprops}, of the two components are within a few \kms\ of one another and are thus consistent with bound Keplerian orbits. RVs were not used in selecting the wide binaries, so this validates their status as genuine binaries.  

\begin{table*}
\caption{Observational Properties of Wide Binary Systems}
\label{tab:obsprops}
\addtolength{\tabcolsep}{-2pt}
\begin{tabular}{lllllllllllll}
\hline\hline
Gaia & RA & DEC & Name & RV & $\sigma$RV & SNR & $\varpi$ & PMRA & PMDEC & $G$ & $BP-RP$ & T$_{\mathrm{eff,phot}}$ \\
 & $\mathrm{{}^{\circ}}$ & $\mathrm{{}^{\circ}}$ &  &\kms  & \kms & pixel$^{-1}$ & $\mathrm{mas}$ & $\mathrm{mas\,yr^{-1}}$ & $\mathrm{mas\,yr^{-1}}$ & $\mathrm{mag}$ & $\mathrm{mag}$ & $\mathrm{K}$ \\
 \hline
1019003329101872896 & 140.6659 & 50.6039 & WB01A & 4.74 & 0.08 & 87 & 15.05 & 52.62 & 9.94 & 8.95 & 0.93 & 5604 \\
1019003226022657920 & 140.6570 & 50.6038 & WB01B & 4.45 & 0.07 & 83 & 15.02 & 55.93 & 10.34 & 8.82 & 0.91 & 5663 \\
1448493530351691520 & 203.6017 & 26.2772 & WB02A & -4.63 & 0.21 & 121 & 7.73 & 9.77 & 1.42 & 8.87 & 0.63 & 6460 \\
1448493427272476288 & 203.5992 & 26.2761 & WB02B & -4.51 & 0.22 & 142 & 7.78 & 9.74 & 1.02 & 8.60 & 0.62 & 6360 \\
219605599154126976 & 57.9389 & 34.8895 & WB03A & 12.42 & 0.36 & 98 & 7.26 & -9.69 & -14.11 & 9.63 & 0.68 & 6440 \\
219593745044391552 & 57.9426 & 34.8830 & WB03B & 12.60 & 0.28 & 111 & 7.21 & -9.70 & -12.45 & 9.79 & 0.70 & 6461 \\
232899966044906496 & 63.9372 & 45.3918 & WB04A & 65.15 & 0.10 & 121 & 12.91 & 130.58 & -205.86 & 8.68 & 0.79 & 5828 \\
232899966044905472 & 63.9385 & 45.3929 & WB04B & 65.14 & 0.10 & 98 & 12.95 & 126.08 & -203.09 & 8.58 & 0.78 & 5886 \\
238164255921243776 & 53.8536 & 42.3046 & WB05A & 17.58 & 0.11 & 87 & 6.07 & 1.59 & -39.61 & 10.06 & 0.82 & 6027 \\
238163534366737792 & 53.8673 & 42.3006 & WB05B & 17.11 & 0.11 & 78 & 5.96 & 2.78 & -39.37 & 9.86 & 0.83 & 5995 \\
2493516351151864960 & 36.2245 & -2.1121 & WB06A & 26.25 & 0.17 & 114 & 8.17 & 70.17 & -13.37 & 8.98 & 0.64 & 6302 \\
2493516351151865088 & 36.2224 & -2.1122 & WB06B & 26.51 & 0.15 & 124 & 8.15 & 69.20 & -14.46 & 9.21 & 0.66 & 6251 \\
2565584837226776448 & 24.4075 & 7.1462 & WB07A & 23.71 & 0.08 & 74 & 14.58 & 79.89 & -79.66 & 9.58 & 1.01 & 5247 \\
2565584802867037696 & 24.4184 & 7.1488 & WB07B & 24.59 & 0.07 & 68 & 14.48 & 80.61 & -78.94 & 9.55 & 1.00 & 5250 \\
2572433351559023616 & 26.0631 & 9.4849 & WB08A & 7.11 & 0.29 & 140 & 13.80 & 139.94 & -68.27 & 7.89 & 0.62 & 6513 \\
2572433347264096768 & 26.0639 & 9.4838 & WB08B & 7.79 & 0.43 & 97 & 13.72 & 142.94 & -67.96 & 7.71 & 0.59 & 6682 \\
2573278051366910336 & 25.3299 & 10.1139 & WB09A & 18.91 & 0.13 & 118 & 12.18 & 161.36 & 34.43 & 8.68 & 0.71 & 5908 \\
2573278120086386432 & 25.3244 & 10.1179 & WB09B & 18.95 & 0.14 & 120 & 12.25 & 159.97 & 35.27 & 8.82 & 0.72 & 6091 \\
271977330850893568 & 65.6146 & 51.8143 & WB10A & 14.94 & 0.15 & 92 & 8.85 & 23.35 & -47.27 & 9.93 & 0.78 & 6242 \\
271977330850895488 & 65.6106 & 51.8089 & WB10B & 15.59 & 0.11 & 80 & 8.90 & 21.34 & -48.69 & 9.67 & 0.74 & 6025 \\
3097066080667487488 & 125.7440 & 7.6303 & WB11A & -15.38 & 0.09 & 97 & 19.62 & 13.90 & -19.06 & 8.98 & 0.96 & 5479 \\
3097066080667486592 & 125.7474 & 7.6308 & WB11B & -14.76 & 0.09 & 97 & 19.66 & 16.17 & -21.31 & 9.02 & 0.97 & 5355 \\
3170300942420466176 & 112.4253 & 18.2757 & WB12A & -26.74 & 0.12 & 110 & 8.07 & -8.96 & -31.39 & 9.28 & 0.77 & 5913 \\
3170394607068638336 & 112.4483 & 18.2795 & WB12B & -26.52 & 0.11 & 105 & 8.13 & -8.81 & -31.62 & 9.50 & 0.77 & 5916 \\
3230677870385455232 & 69.3588 & 0.5747 & WB13A & 39.18 & 0.13 & 129 & 15.60 & 15.95 & 12.16 & 7.36 & 0.74 & 5864 \\
3230677565443833088 & 69.3614 & 0.5532 & WB13B & 39.15 & 0.14 & 134 & 15.59 & 16.65 & 11.60 & 7.34 & 0.74 & 5865 \\
3288572968680438912 & 73.5689 & 7.3680 & WB14A & 47.57 & 0.08 & 101 & 33.78 & 246.13 & -197.63 & 8.11 & 1.07 & 5193 \\
3288572968680438528 & 73.5704 & 7.3722 & WB14B & 47.41 & 0.08 & 101 & 33.75 & 248.06 & -202.01 & 7.96 & 1.04 & 5226 \\
3391840612589045632 & 77.5610 & 13.9951 & WB15A & 37.16 & 1.45 & 154 & 9.68 & 13.57 & -10.18 & 8.48 & 0.62 & 6519 \\
3391840539572707072 & 77.5628 & 13.9880 & WB15B & 37.40 & 0.18 & 133 & 9.49 & 13.32 & -11.56 & 8.73 & 0.66 & 6493 \\
3588936180766441600 & 177.0016 & -8.6279 & WB16A & -19.04 & 0.25 & 71 & 7.26 & 61.62 & -47.01 & 9.44 & 0.64 & 6258 \\
3588936180766441728 & 176.9990 & -8.6263 & WB16B & -18.71 & 0.17 & 60 & 7.32 & 62.40 & -47.03 & 9.67 & 0.66 & 6354 \\
3644886925888351872 & 209.0257 & -4.6167 & WB17A & 8.34 & 0.20 & 46 & 9.01 & -11.39 & 26.99 & 8.96 & 0.67 & 6095 \\
3644886925888352000 & 209.0271 & -4.6159 & WB17B & 7.24 & 0.24 & 67 & 8.91 & -10.73 & 26.60 & 8.91 & 0.66 & 6265 \\
3890860183966486656 & 156.7826 & 18.0623 & WB18A & 12.03 & 0.08 & 88 & 17.48 & -124.03 & -105.54 & 9.21 & 0.98 & 5347 \\
3890860179670959104 & 156.7845 & 18.0619 & WB18B & 11.86 & 0.08 & 93 & 17.48 & -124.84 & -112.56 & 9.17 & 0.96 & 5404 \\
3975129194660883328 & 178.6316 & 19.4112 & WB19A & 6.52 & 0.10 & 126 & 25.24 & -450.50 & -16.55 & 8.03 & 0.86 & 5739 \\
3975223065466473216 & 178.6441 & 19.4278 & WB19B & 6.31 & 0.09 & 125 & 25.25 & -450.60 & -15.50 & 8.22 & 0.90 & 5607 \\
4024887730814401280 & 174.7115 & 32.6420 & WB20A & 22.88 & 0.15 & 113 & 7.51 & -94.48 & 44.03 & 9.89 & 0.70 & 6195 \\
4024886425144354816 & 174.6790 & 32.6261 & WB20B & 20.57 & 0.14 & 114 & 7.56 & -95.34 & 43.59 & 9.78 & 0.69 & 6283 \\
440947391590004096 & 46.0106 & 52.5151 & WB21A & -38.83 & 0.16 & 98 & 7.10 & 23.25 & 25.53 & 9.50 & 0.78 & 5947 \\
440959142620525568 & 46.0079 & 52.5165 & WB21B & -39.66 & 0.16 & 80 & 7.07 & 23.64 & 24.43 & 9.73 & 0.78 & 6040 \\
478240661338191360 & 75.8897 & 63.0873 & WB22A & -25.49 & 0.09 & 123 & 31.08 & 101.97 & 316.19 & 7.44 & 0.84 & 5514 \\
478240661338195328 & 75.8793 & 63.0795 & WB22B & -24.84 & 0.10 & 136 & 31.04 & 105.90 & 311.37 & 7.73 & 0.89 & 5687 \\
692119656035933568 & 137.0992 & 27.5357 & WB23A & 31.05 & 0.12 & 111 & 20.45 & -53.24 & 71.66 & 8.11 & 0.74 & 5996 \\
692120029700390912 & 137.1129 & 27.5434 & WB23B & 31.31 & 0.13 & 116 & 20.36 & -51.82 & 73.52 & 8.13 & 0.74 & 5974 \\
736174028943041920 & 159.8656 & 31.7048 & WB24A & 9.22 & 0.10 & 97 & 14.48 & -110.90 & -36.06 & 9.20 & 0.89 & 5632 \\
736173925863826944 & 159.8621 & 31.7008 & WB24B & 8.65 & 0.09 & 83 & 14.43 & -112.71 & -36.51 & 9.19 & 0.90 & 5605 \\
914241517609344128 & 126.7051 & 39.0131 & WB25A & -3.33 & 0.20 & 114 & 11.64 & 27.13 & 5.81 & 8.69 & 0.66 & 6430 \\
914244399532441472 & 126.6368 & 39.0493 & WB25B & -3.09 & 0.16 & 127 & 11.72 & 27.62 & 6.36 & 8.93 & 0.71 & 6232 \\
\hline \hline
\end{tabular}
\raggedright
NOTE: The \gaia\ DR2 source identifier of each star is given in column~1 with sky coordinates in columns~2 and 3. The associated wide binary component of each star is listed in column~4. The radial velocity and its uncertainty measured in our optical spectra are given columns~5 and 6, respectively. The signal-to-noise ratios (SNR), measured in at the continuum level at $\sim$5350\AA, of our optical spectra are listed in column 7.  The parallax and proper motion in RA and DEC are given in columns 8, 9, and 10, respectively. The $G$ band magnitude and $(BP-RP)$ colour are given in columns 11 and 12, respectively. Finally the \gaia\ derived photometric temperatures \citep{Andrae2018} are given in the last column.
\end{table*}

The final reduced spectra have wavelength coverage  $\sim$3500-10000~\AA\ over $\sim$60 echelle orders with some inter-order gaps, particularly in the redder wavelengths. In Table~\ref{tab:obsprops}, we report the basic observational properties (i.e. \gaia\ DR2 source identified numbers, sky positions, parallaxes, proper motions, radial velocities, photometry) of our sample. We also report the photometric \teff\ provided by \gaia\ \citep[their \textsc{teff\_val} column,][]{Andrae2018} in order to compare to the temperatures derived spectroscopically in this work. The typical uncertainty on the photometric \teff\ values derived from \gaia\ are on the order of $\sigma$\teff\ $\sim$150~K.

 We primarily obtained high signal-to-noise ratio (SNR $>$ 60~pixel$^{-1}$) for each star in the \numpair\ co-moving pair in order to precisely quantify their chemical abundance pattern. We note that the typical (mean) SNR is $\sim$ 105~pixel$^{-1}$ ensuring that we can obtain high fidelity chemical abundance estimates. In Fig.~\ref{fig:spectra} we show sample spectra of 4 pairs. It is interesting to already note that the spectra of the various pairs look remarkably similar.

\section{Stellar Parameter and Abundance Analysis} \label{sec:SP}
Stellar parameters were determined in an automatic fashion under the standard Fe excitation-ionization balance technique using the `param' module of the Brussels Automatic Code for Characterizing High accUracy Spectra \citep[BACCHUS,][]{Masseron2016} code. Similar to \cite{Hawkins2018b}, we used the version of BACCHUS which includes the MARCS model atmosphere grid \citep{Gustafsson2008}, along with TURBOSPECTRUM \citep{Alvarez1998, Plez2012}, which is used to generate synthetic spectra under the assumption of Local Thermodynamic Equilibrium (LTE). The atomic data (line list) are taken from the fifth version of the Gaia-ESO linelist \citep{Heiter2019}. Molecular species were also included. The molecular species added include:  CH \citep{Masseron2014}, and CN, NH, OH, MgH and  C$_{2}$(T. Masseron, private communication). SiH molecules are adopted from the Kurucz linelists\footnote{http://kurucz.harvard.edu/linelists/linesmol/} and those from TiO, ZrO, FeH, CaH from B. Plez (private communication) are also included. We note that hyperfine structure splitting is included for Sc I, V I Mn I, Co I, Cu I, Ba II, Eu II, La II, Pr II, Nd II, Sm II \citep{Heiter2019}. The synthetic spectra produced using the above procedure are then compared via $\chi^2$ minimization to the observed spectra. We note here that instrument, rotational, and macroturbulent broadening are also included during the spectral synthesis and derived by ensuring that the abundance determined from the $\chi^2$ minimization matches the abundances determined using the core of the line \citep{Masseron2016}. The line selection for each element was done in the same way as in \cite{Hawkins2015}.

Under the standard Fe excitation-ionization balance procedure we derived the \teff\  by ensuring that there is no correlation between the abundance of Fe and the excitation potential of the lines being used. On the other hand, \logg\ is derived by forcing no significant offset between the abundance of neutral Fe (\ion{Fe}{I}) and that of singly ionized Fe (\ion{Fe}{II}). Further, the microturbulent velocity parameter \vmicro\ is determined by ensuring there to be no correlation between the abundance of Fe and the reduced equivalent width (REW, defined as equivalent width divided by the wavelength of the line). For this procedure we used up to 100~\ion{Fe}{I} lines and 20~\ion{Fe}{II} lines. Individual abundances for \numel\ elements across the light/odd-Z (Li, C, Na, Al, Sc, V), $\alpha$ (Mg, Si, Ca, Ti), Fe-peak (Cr, Mn, Fe, Co, Ni, Zn), and neutron capture (Sr, Y, Zr, Ba, La, Nd, Eu) families were derived using the `abund' module within BACCHUS. This module derives abundances by first fixing the stellar atmospheric parameters to those derived as described using Fe excitation-ionization balance  and  then synthesizing spectra with different values of [X/H]. The reported [X/H] abundance was determined using a $\chi^2$ minimization between these synthetic and observed spectra. For more details about BACCHUS, we refer the reader to Section~2.2 of \cite{Hawkins2015}. We note that the solar abundances from \cite{Asplund2005} are assumed. 
The total internal uncertainty in the derived abundances is important to quantify in order to determine with what precision we can conclude the chemical homogeneity of wide binary systems. In order to derive an estimate of the total internal uncertainty in stellar abundances, we follow \cite{Hawkins2016b} where we first quantify the representative sensitivity of each of the chemical abundances to the uncertainty in the stellar atmospheric parameters (\teff, \logg, \vmicro). This is effectively equivalent to propagating the uncertainty in the stellar parameters through to the chemical abundances. The sensitivity of the abundance ratios, [X/H], due to uncertainties in the  \teff, \logg, and \vmicro\ are then added in quadrature with standard error in the mean of the individual absorption features used to determine the abundance \citep[e.g.][]{Desidera2004, Desidera2006, Yong2013, Roederer2014, Hawkins2016b, Lucey2019}. We note here that, in principle, this method for estimating the uncertainty in the stellar abundances neglect the covariances between the stellar parameters and treats them independently \citep[e.g. see][for a longer discussion on this]{McWilliam1995}. In Table~\ref{tab:uncert}, we tabulate the typical (conservative) sensitivities  (i.e. the median difference in [X/H]) due to an uncertainty in \teff\ of 100~K, in \logg\ of 0.25~dex, and \vmicro\ of 0.10~\kms. These typical uncertainties are computed as the mean of the \teff, \logg, \feh, and \vmicro\ uncertainty which is an output of the `param' module within BACCHUS. For this calculation, we choose to compute the sensitivities for 8 stars across 4 binary systems that span our \teff-\logg-\feh\ parameter range. The median difference in [X/H] as a result of perturbing the stellar parameters by their uncertainties can be found in Table~\ref{tab:uncert}. 


\begin{table}
\caption{Stellar abundance sensitivities to the uncertainty in the stellar parameters }
\label{tab:uncert}
\centering
\begin{tabular}{c c c c}
\hline\hline
$\Delta$[X/H] & $\Delta$\teff\ &$\Delta$\logg\ & $\Delta$\vmicro\  \\
 & ($\pm$100~K)&  ($\pm$0.25~dex) &  ($\pm$0.10~\kms)\\
 \hline
Li&$\pm$0.07&$\pm$0.00&$\mp$0.00\\
C&$\pm$0.11&$\mp$0.02&$\mp$0.00\\
Na&$\pm$0.06&$\mp$0.03&$\mp$0.00\\
Mg&$\pm$0.11&$\mp$0.10&$\mp$0.01\\
Al&$\pm$0.04&$\pm$0.00&$\mp$0.00\\
Si&$\pm$0.03&$\pm$0.02&$\mp$0.02\\
Ca&$\pm$0.06&$\mp$0.02&$\mp$0.02\\
Sc&$\pm$0.03&$\pm$0.07&$\mp$0.03\\
Ti&$\pm$0.10&$\pm$0.01&$\mp$0.03\\
V&$\pm$0.12&$\pm$0.01&$\mp$0.00\\
Cr&$\pm$0.08&$\mp$0.01&$\mp$0.02\\
Mn&$\pm$0.09&$\mp$0.01&$\mp$0.02\\
Fe&$\pm$0.06&$\pm$0.01&$\mp$0.02\\
Co&$\pm$0.08&$\pm$0.03&$\pm$0.00\\
Ni&$\pm$0.07&$\mp$0.01&$\mp$0.02\\
Cu&$\pm$0.07&$\pm$0.01&$\mp$0.01\\
Zn&$\pm$0.03&$\pm$0.02&$\mp$0.01\\
Sr&$\pm$0.10&$\pm$0.03&$\pm$0.01\\
Y&$\pm$0.02&$\pm$0.07&$\mp$0.03\\
Zr&$\pm$0.05&$\pm$0.05&$\mp$0.00\\
Ba&$\pm$0.04&$\pm$0.01&$\mp$0.05\\
La&$\pm$0.05&$\pm$0.07&$\mp$0.01\\
Nd&$\pm$0.04&$\pm$0.06&$\mp$0.00\\
Eu&$\pm$0.01&$\pm$0.09&$\mp$0.01\\
 \hline \hline
\end{tabular}
\raggedright
\\ NOTE: The change in [X/H] abundance (denoted in column~1) when the  \teff\ is perturbed by $\pm$100~K (column~2), \logg\ is perturbed by $\pm$0.25~dex (column~3), \vmicro\ is perturbed by $\pm$0.10~\kms (column~4). Total uncertainties are obtained by adding these in quadrature with the standard error in the line-by-line abundances.
\end{table}

\section{Results and Discussion} \label{sec:results}
In this section, we present the results of our stellar parameter and abundance analysis and critically focus on the difference in chemical abundance ratios, i.e. $\Delta$[X/H] and $\Delta$[X/Fe], between the two stars in the \numpair\ wide binary systems. We also intermix these results with a discussion placing these results in the context of recent studies on the homogeneity of wide binary systems. We start by presenting the stellar parameters for each of the \numstar\ observed stars in the \numpair\ binary systems in section~\ref{subsec:SP}. We then move on to discuss the differences in the [X/H, Fe] abundance ratios for wide binaries compared to random pairings of stars in light and odd-Z elements (section~\ref{sub:light}), $\alpha$ elements (section~\ref{sub:alpha}),  Fe-peak elements (section~\ref{sub:Fepeak}), and neutron capture elements (section~\ref{sub:ncap}).


\begin{table*}
\caption{Stellar Parameter and Chemical Abundance Ratios of Observed  Wide Binary Systems}

\begin{tabular}{ccccccccccccc}
\hline\hline
Source ID & Name&\teff & $\sigma$\teff & \logg & $\sigma$\logg & [Fe/H] & $\sigma$[Fe/H] & \vmicro & $\sigma$\vmicro & [Si/H] & $\sigma$[Si/H] & ... \\
 & & (K) & (K)&(dex)&(dex)&(dex)&(dex)&(\kms)&(\kms)&(dex)&(dex)&\\
 \hline
1019003329101872896 & WB01A & 5604 & 24 & 4.62 & 0.07 & 0.36 & 0.01 & 0.87 & 0.05 & 0.41 & 0.03 & ... \\
1019003226022657920 & WB01B & 5663 & 21 & 4.67 & 0.12 & 0.37 & 0.01 & 1.03 & 0.05 & 0.34 & 0.03 & ... \\
1448493530351691520 & WB02A & 6460 & 54 & 3.96 & 0.45 & -0.11 & 0.01 & 1.47 & 0.08 & -0.21 & 0.02 & ... \\
1448493427272476288 & WB02B & 6360 & 95 & 3.94 & 0.19 & -0.21 & 0.01 & 1.62 & 0.09 & -0.22 & 0.04 & ... \\
219605599154126976 & WB03A & 6440 & 38 & 4.36 & 0.34 & -0.35 & 0.02 & 1.09 & 0.11 & -0.29 & 0.05 & ... \\
219593745044391552 & WB03B & 6461 & 64 & 4.56 & 0.68 & -0.34 & 0.01 & 1.47 & 0.14 & -0.34 & 0.05 & ... \\
232899966044906496 & WB04A & 5828 & 31 & 4.35 & 0.20 & -0.00 & 0.01 & 1.17 & 0.05 & 0.03 & 0.03 & ... \\
232899966044905472 & WB04B & 5886 & 29 & 4.44 & 0.29 & 0.04 & 0.01 & 1.16 & 0.05 & 0.05 & 0.02 & ... \\
238164255921243776 & WB05A & 6027 & 29 & 4.29 & 0.31 & 0.20 & 0.01 & 1.25 & 0.04 & 0.18 & 0.03 & ... \\
238163534366737792 & WB05B & 5995 & 68 & 4.32 & 0.44 & 0.09 & 0.01 & 1.32 & 0.05 & 0.14 & 0.03 & ... \\
2493516351151864960 & WB06A & 6302 & 108 & 4.22 & 0.26 & -0.20 & 0.01 & 1.54 & 0.07 & -0.19 & 0.03 & ... \\
2493516351151865088 & WB06B & 6251 & 52 & 4.30 & 0.49 & -0.19 & 0.01 & 1.36 & 0.06 & -0.17 & 0.03 & ... \\
2565584837226776448 & WB07A & 5247 & 22 & 4.67 & 0.20 & 0.14 & 0.01 & 0.70 & 0.04 & 0.20 & 0.04 & ... \\
2565584802867037696 & WB07B & 5250 & 21 & 4.56 & 0.23 & 0.13 & 0.01 & 0.88 & 0.04 & 0.19 & 0.04 & ... \\
2572433351559023616 & WB08A & 6513 & 44 & 4.07 & 0.85 & -0.06 & 0.01 & 1.37 & 0.07 & -0.07 & 0.04 & ... \\
2572433347264096768 & WB08B & 6682 & 15 & 4.39 & 0.49 & -0.07 & 0.02 & 1.35 & 0.12 & 0.01 & 0.04 & ... \\
2573278051366910336 & WB09A & 5908 & 61 & 4.31 & 0.28 & -0.28 & 0.01 & 1.27 & 0.06 & -0.23 & 0.03 & ... \\
2573278120086386432 & WB09B & 6091 & 39 & 4.31 & 0.39 & -0.16 & 0.01 & 1.24 & 0.06 & -0.19 & 0.02 & ... \\
271977330850893568 & WB10A & 6242 & 32 & 4.83 & 0.28 & -0.01 & 0.01 & 1.22 & 0.06 & -0.08 & 0.04 & ... \\
271977330850895488 & WB10B & 6025 & 17 & 4.80 & 0.28 & 0.02 & 0.01 & 1.02 & 0.04 & -0.09 & 0.04 & ... \\
3097066080667487488 & WB11A & 5479 & 65 & 4.78 & 0.33 & -0.01 & 0.01 & 1.52 & 0.05 & -0.05 & 0.03 & ... \\
3097066080667486592 & WB11B & 5355 & 57 & 4.75 & 0.01 & -0.06 & 0.01 & 1.20 & 0.05 & -0.02 & 0.03 & ... \\
3170300942420466176 & WB12A & 5913 & 24 & 4.21 & 0.17 & -0.20 & 0.01 & 1.13 & 0.05 & -0.20 & 0.02 & ... \\
3170394607068638336 & WB12B & 5916 & 14 & 4.27 & 0.25 & -0.16 & 0.01 & 1.04 & 0.04 & -0.22 & 0.02 & ... \\
3230677870385455232 & WB13A & 5864 & 69 & 3.92 & 0.10 & -0.30 & 0.01 & 1.37 & 0.06 & -0.22 & 0.03 & ... \\
3230677565443833088 & WB13B & 5865 & 24 & 3.79 & 0.21 & -0.33 & 0.01 & 1.31 & 0.08 & -0.28 & 0.03 & ... \\
3288572968680438912 & WB14A & 5193 & 68 & 4.50 & 0.44 & 0.07 & 0.01 & 1.22 & 0.04 & 0.01 & 0.03 & ... \\
3288572968680438528 & WB14B & 5226 & 38 & 4.65 & 0.07 & 0.06 & 0.01 & 1.13 & 0.04 & 0.09 & 0.03 & ... \\
3391840612589045632 & WB15A & 6519 & 65 & 4.38 & 0.26 & 0.13 & 0.01 & 1.57 & 0.09 & 0.10 & 0.02 & ... \\
3391840539572707072 & WB15B & 6493 & 69 & 4.34 & 0.31 & 0.12 & 0.01 & 1.55 & 0.08 & 0.09 & 0.02 & ... \\
3588936180766441600 & WB16A & 6258 & 47 & 4.13 & 0.26 & -0.30 & 0.01 & 1.39 & 0.09 & -0.25 & 0.03 & ... \\
3588936180766441728 & WB16B & 6354 & 51 & 4.44 & 0.49 & -0.17 & 0.01 & 1.31 & 0.10 & -0.21 & 0.03 & ... \\
3644886925888351872 & WB17A & 6095 & 106 & 4.06 & 0.34 & -0.00 & 0.02 & 1.55 & 0.09 & 0.01 & 0.02 & ... \\
3644886925888352000 & WB17B & 6265 & 83 & 4.15 & 0.34 & 0.01 & 0.01 & 1.71 & 0.07 & -0.02 & 0.04 & ... \\
3890860183966486656 & WB18A & 5347 & 64 & 4.60 & 0.32 & 0.07 & 0.01 & 1.10 & 0.04 & 0.06 & 0.03 & ... \\
3890860179670959104 & WB18B & 5404 & 17 & 4.64 & 0.32 & 0.05 & 0.01 & 1.05 & 0.03 & 0.05 & 0.03 & ... \\
3975129194660883328 & WB19A & 5739 & 25 & 4.69 & 0.23 & -0.07 & 0.01 & 1.00 & 0.05 & -0.10 & 0.02 & ... \\
3975223065466473216 & WB19B & 5607 & 14 & 4.75 & 0.12 & -0.04 & 0.01 & 0.76 & 0.05 & -0.09 & 0.03 & ... \\
4024887730814401280 & WB20A & 6195 & 25 & 4.47 & 0.30 & -0.16 & 0.01 & 1.10 & 0.06 & -0.22 & 0.02 & ... \\
4024886425144354816 & WB20B & 6283 & 26 & 4.55 & 0.30 & -0.14 & 0.01 & 1.19 & 0.06 & -0.18 & 0.03 & ... \\
440947391590004096 & WB21A & 5947 & 129 & 4.16 & 0.23 & -0.66 & 0.01 & 1.59 & 0.15 & -0.59 & 0.03 & ... \\
440959142620525568 & WB21B & 6040 & 45 & 4.21 & 0.31 & -0.58 & 0.01 & 1.20 & 0.13 & -0.54 & 0.03 & ... \\
478240661338191360 & WB22A & 5514 & 26 & 4.64 & 0.21 & -0.22 & 0.01 & 1.01 & 0.05 & -0.19 & 0.03 & ... \\
478240661338195328 & WB22B & 5687 & 25 & 4.57 & 0.38 & -0.21 & 0.01 & 1.06 & 0.04 & -0.19 & 0.02 & ... \\
692119656035933568 & WB23A & 5996 & 30 & 4.62 & 0.22 & -0.26 & 0.01 & 1.07 & 0.06 & -0.31 & 0.02 & ... \\
692120029700390912 & WB23B & 5974 & 68 & 4.51 & 0.33 & -0.28 & 0.01 & 1.30 & 0.05 & -0.33 & 0.03 & ... \\
736174028943041920 & WB24A & 5632 & 45 & 4.63 & 0.35 & 0.17 & 0.01 & 1.13 & 0.04 & 0.13 & 0.03 & ... \\
736173925863826944 & WB24B & 5605 & 50 & 4.63 & 0.38 & 0.19 & 0.01 & 1.17 & 0.04 & 0.17 & 0.03 & ... \\
914241517609344128 & WB25A & 6430 & 37 & 4.59 & 0.23 & -0.04 & 0.01 & 1.28 & 0.07 & -0.12 & 0.03 & ... \\
914244399532441472 & WB25B & 6232 & 34 & 4.64 & 0.20 & -0.10 & 0.01 & 1.29 & 0.09 & -0.12 & 0.03 & ... \\
\hline\hline
\end{tabular}
\label{tab:results}
\raggedright
NOTE: This is a subsample of the spectroscopically derived stellar parameters (\teff, \logg, \feh, \vmicro) and the chemical abundances [X/H] for the \numstar\ stars in our sample. The full table will be provided in the online material. The \gaia\ DR2 source identifier and the wide binary name of each star is given in columns~1 and 2. The stellar parameters and their uncertainties (\teff, $\sigma$\teff, \logg, $\sigma$\logg, \feh (where [Fe/H] is measured by [Fe I/H]), $\sigma$\feh, \vmicro, $\sigma$\vmicro) are found in columns 3-10, respectively. The chemical abundance ratio for [Si/H] is found in column 11 and its uncertainty in column 12. We note that this uncertainty is determined as the dispersion in the [Si/H] over all lines used to derive [Si/H] divided by the square root of the number of lines used.  
\end{table*}

\subsection{Stellar Atmospheric Parameters}\label{subsec:SP}
The stellar atmospheric parameters and chemical abundances, denoted [X/H], can be found in Table~\ref{tab:results}. More specifically, the derived stellar parameters, and their uncertainties are reported in the first eight columns after the identifiers in Table~\ref{tab:results}. Typical errors in \teff, \logg, \feh, and \vmicro\ are approximately $\sim$40~K, 0.25~dex, 0.01~dex (line-by-line), and 0.06~\kms, respectively. Additionally, the chemical abundances (reported as [X/H]) for \numel\ elemental species for each star in our sample can also be found in Table~\ref{tab:results}\footnote{The full table will be provided as an online table. Here we show a cut out of the full table for reference.}. These are determined by taking the median of the [X/H] abundances in `clean' absorption features found by the BACCHUS `abund' module. 

\begin{figure*}
	 \includegraphics[width=2\columnwidth]{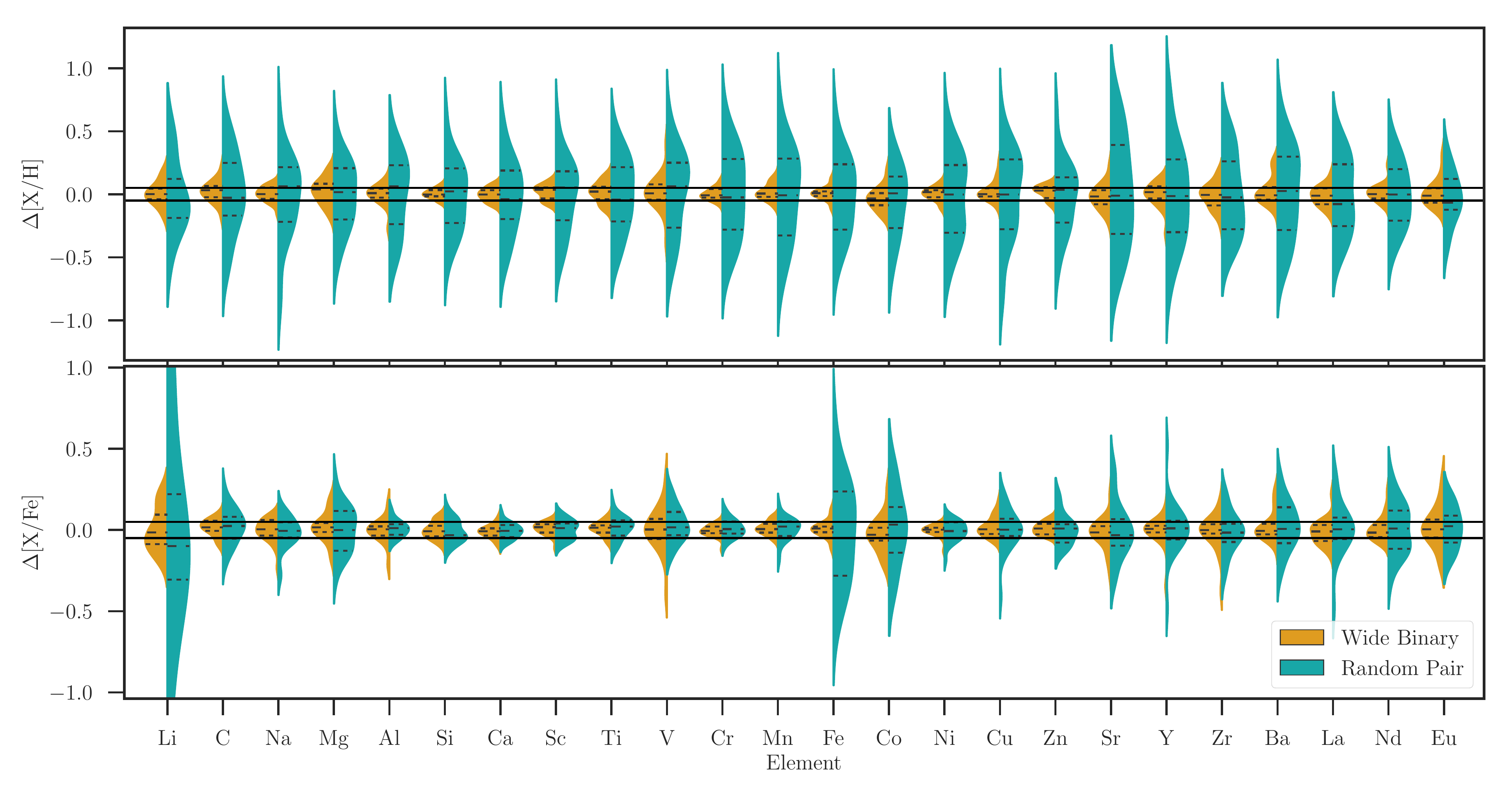}
	\caption{Top: Violin diagram showing the distribution of the difference in [X/H] for the \numel\ reported elemental species between the two components of the \numpair\ wide binary pairs (orange). Also shown is the distribution of the difference in [X/H] between each star and the closest star on the color-magnitude diagram (Fig.~\ref{fig:CMD}), which is not its companion (cyan).  Bottom: The same as the top panel but now showing the difference in [X/Fe] instead of [X/H]. For reference, dashed lines denote the (inner, outer and median) quartiles and solid lines in both panels are shown at $\Delta$[X/Fe]~=~$\pm$0.05 and $\Delta$[X/H]~=~$\pm$0.05~dex. For reference, in the bottom panel, Fe represents the difference in the [Fe/H].  }
	\label{fig:XHFediff}
\end{figure*}

The uncertainties of the reported [X/H] abundances in that table are derived by taking the standard deviation in the line-by-line [X/H] abundances and dividing by the square root of the number of lines used (i.e. the standard error in the mean). Where only one line is able to be measured the uncertainty is conservatively assumed to be $\pm$0.10~dex.  The {\it total abundance uncertainty} is determined by adding, in quadrature, the uncertainty in mean [X/H], which is reported in Table~\ref{tab:results}, along with each of the typical sensitivities of the abundance ratio with respect to the  stellar parameters, reported in Table~\ref{tab:uncert}. The median total abundance uncertainty across all elements is on the order of $\sim\pm$0.08~dex. 

In practice, for both chemical tagging and characterization of exotic (M-dwarf, white-dwarf, etc.) stars using wide binaries to work, the difference in [X/H], or conversely [X/Fe], between the two stars in the pair must be consistent with zero. In this case, both stars in the wide binary pair would be chemically identical. Therefore, the distribution of difference for each chemical element ratio (with the $\Delta$[X/H] in the top panel and $\Delta$[X/Fe] in the bottom panel) between the wide binary pairs (orange) in this work are shown as a violin diagram in Fig.~\ref{fig:XHFediff}. For each element, we take the [X/H, Fe] ratio of component A and subtract it from the [X/H, Fe] ratio of component B. Stars in each pair were randomly assigned an `A' or `B' label. For reference, we also show in cyan the distribution of the difference in [X/H] (top) and [X/Fe] (bottom) between one star in each pair and the closest star in color-magnitude space that is {\it not its  companion.} This can be thought of as a `random pairing' of stars which also happen to have similar \teff.  We choose only one star per pair to match with a random star to in order to consistently compare 25 random pairs to 25 wide binary pairs. This was done to compare the chemical homogeneity of random pairings of field stars of similar stellar parameters but {\it not} born together to those wide binary systems which are likely born together.  For reproducibility, we have identified the random pair combinations used for this work in Table~\ref{tab:results}. We also note here the key results do not change by using completely random pairs versus those which are random but also close-by in colour-magnitude space.

\begin{figure}
	 \includegraphics[width=1\columnwidth]{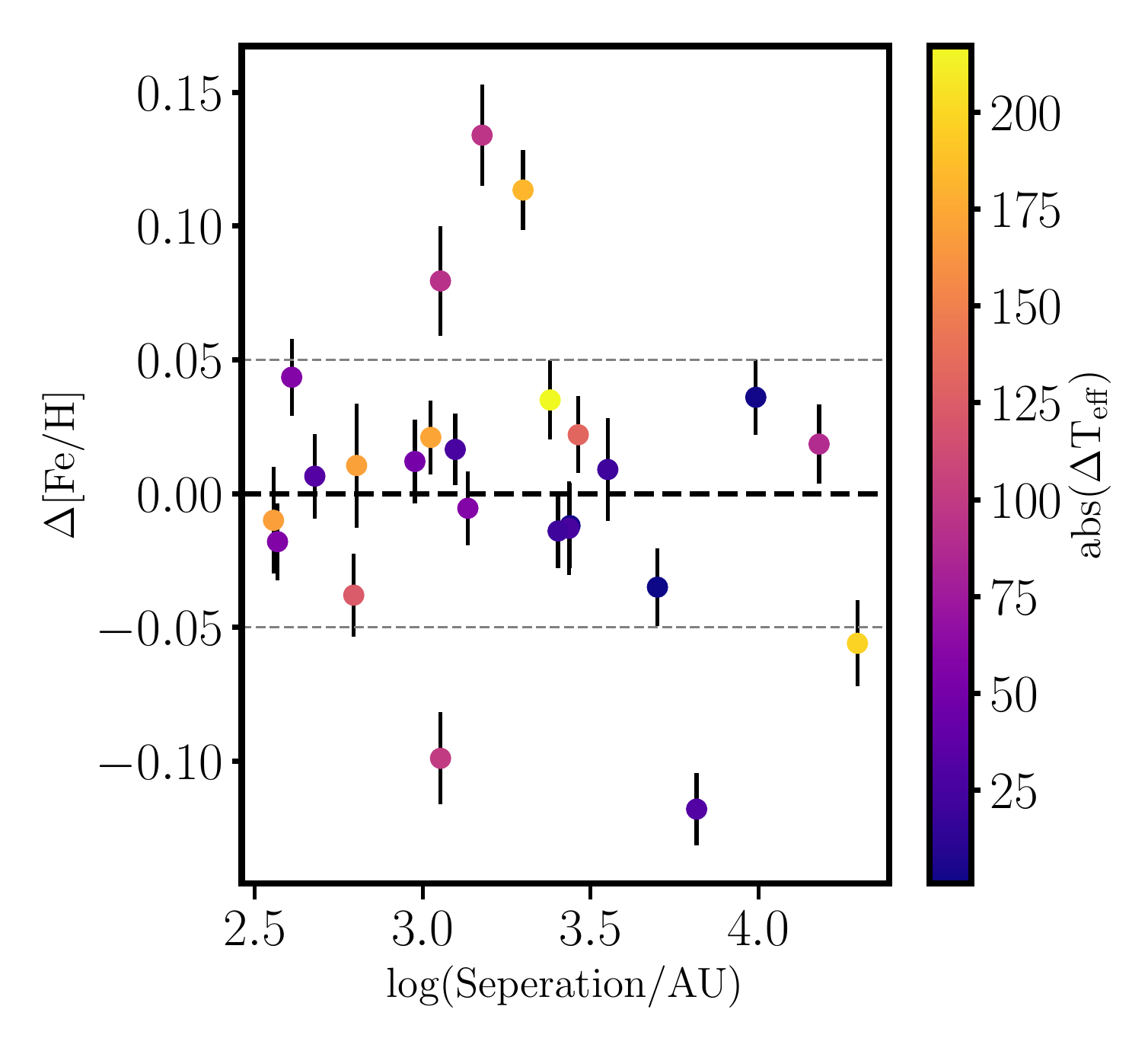}
	\caption{The difference in the metallicity, $\Delta$\feh, of both components of the wide binary as a function of the projected separation between the components. Each pair is colour-coded by the difference in the \teff\ between the stars of the pair.   } 
	\label{fig:sep}
\end{figure}

Focusing first on the wide binaries, in Fig.~\ref{fig:XHFediff} we find that the distribution in $\Delta$\feh\ is centered at $\Delta$\feh\ $\sim$0.00 dex with a dispersion of 0.05~dex. We note however, that there is a component (which accounts for 80\% of the sample, 20/25 systems)  that is chemically homogeneous, with a median $\Delta$\feh\ $\sim$0.01~dex and a dispersion of 0.02~dex and a second component (20\% of the sample or 5/20 systems) of chemically similar, but not homogeneous, wide binaries with a median $\Delta$\feh\ $\sim$0.11~dex and a dispersion of 0.04~dex. Fig.~\ref{fig:sep} shows the difference in \feh\ between wide binary pairs as a function of their separation. The projected separations are taken from \cite{ElBadry2018a}. This figure indicates that the five systems which have $\Delta$\feh\ $>$ 0.10~dex are not at systematically larger separations compared to those which are chemically alike (with $\Delta$\feh\ $<$ 0.01~dex). In a forthcoming work, we will explore pairs with separation $> 10^4$ AU (Ting, Ji, Hawkins, in preparation). \\ 

This result indicates that the occurrence of wide binaries which have large abundance difference is not a common event. Our results also indicate that wide binary systems are {\it commonly homogeneous to within $\pm$0.02~dex in \feh.} This is consistent with and builds on what has been found in other studies \citep[e.g.][]{Desidera2004, Desidera2006, Andrews2017, Andrews2019}. Interestingly, in a smaller sample of 8 wide binaries, \cite{Simpson2018} found that abundance differences between components of wide binary systems observed in the GALAH survey are much more common. This could be due to the fact these authors compare wide binary pairs which have very different effective temperatures ($\Delta$\teff $>$ 200~K). This is known to induce larger abundance differences \citep{Andrews2019}.

\begin{figure}
	 \includegraphics[width=1\columnwidth]{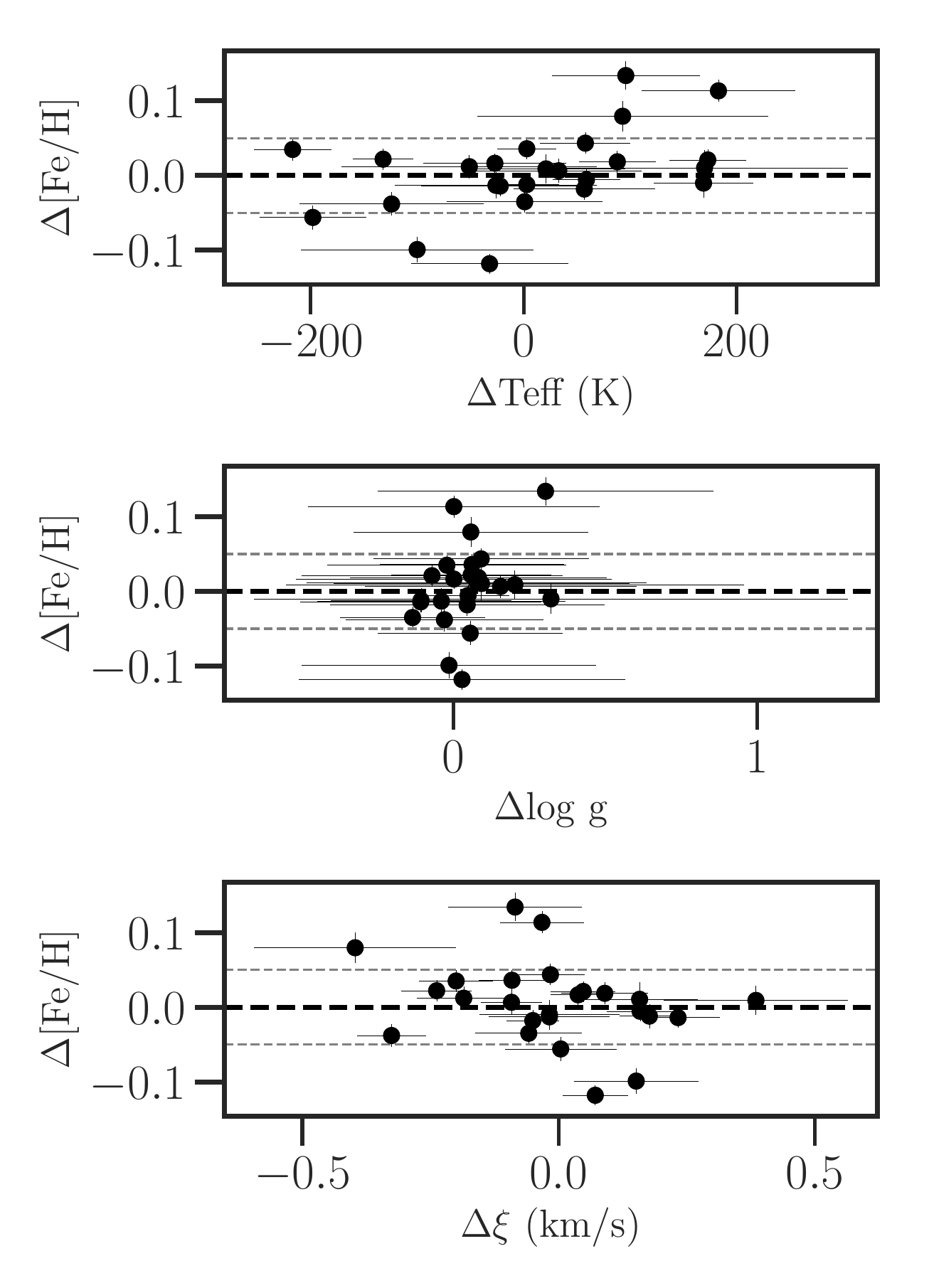}
	\caption{Top:  The difference in \feh\ between wide binary pairs as a function of the difference in their \teff.  Middle: The difference in \feh\ between wide binary pairs as a function of the difference in their \logg. Bottom:  The difference in \feh\  as a function of the difference in their \vmicro.} 
	\label{fig:dFedParam}
\end{figure}

\begin{figure}
	 \includegraphics[width=1\columnwidth]{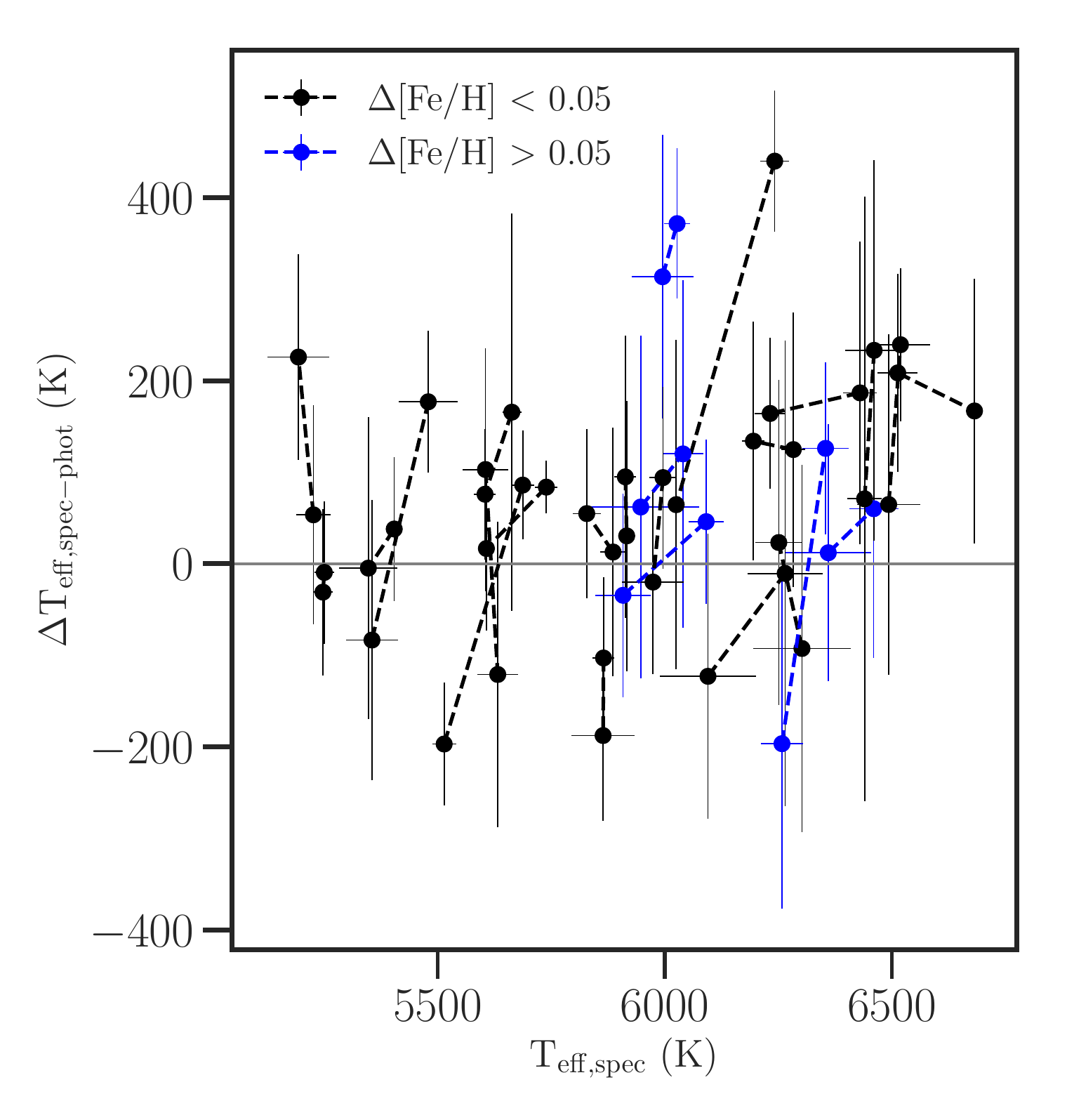}
	\caption{The difference in the photometrically-derived \teff\ and the spectroscopically derived \teff, $\Delta$\teff, as a function of the spectroscopic \teff. Each star in the wide binary pair is connected using a dotted line. Wide binaries with differences in \feh\ less than 0.05~dex are shown in black while those with $\Delta$\feh\ $>$ 0.05~dex are shown in blue. } 
	\label{fig:photTspecT}
\end{figure}

It is possible that there are systematic issues with the \teff\ for the outlier population. Therefore, to ensure that the pairs with $\Delta$\feh\ $> \pm$ 0.10~dex are reliable, in Fig.~\ref{fig:dFedParam}, we show the difference in the \feh\ for each wide binary system as a function of the difference in the $\Delta$\teff\ (top panel), $\Delta$\logg\ (middle panel), $\Delta$\vmicro (bottom panel). We do this as a way to determine whether the outlier wide binary systems, which are different in \feh\, with $\Delta$\feh $>$ 0.10~dex, are a result of the systematics induced by the differences in the stellar parameters between the two stars. Fig.~\ref{fig:dFedParam} show that there are no correlations between the difference in \feh\ and the differences in the remaining stellar parameters. Additionally, in Fig.~\ref{fig:photTspecT} we show the difference in the photometric \teff\ determined from \gaia\ \citep{Andrae2018}, and the spectroscopic \teff\ derived in this work as a function of the spectroscopic \teff. The median offset between the photometric and spectroscopic \teff\ is $\sim$60~K with a dispersion of 130~K. This offset is consistent with other (optical) spectroscopic \teff\ comparisons with photometric \teff\ scales \citep[e.g.][]{ Bergemann2014}. Finally, in  Fig.~\ref{fig:spectra}, in magenta, we show the spectra of one of the wide binary pairs with $\Delta$\feh\ $>$ 0.10~dex compared with spectra from other wide binary pairs which are chemically homogeneous. The spectra between the two stars in WB16 (shown as the magenta solid and dotted lines) are significantly different in the strength of their absorption features, unlike the remaining wide binary pairs. This is to say the spectra for wide binaries that have $\Delta$\feh\ $>$  0.10~dex are visibly different compared to those that are not. 

\label{sub:light}

\subsection{Light/odd-Z elements (Li, C, Na, Al, Sc, V, Cu)}
\begin{figure}
	 \includegraphics[width=1\columnwidth]{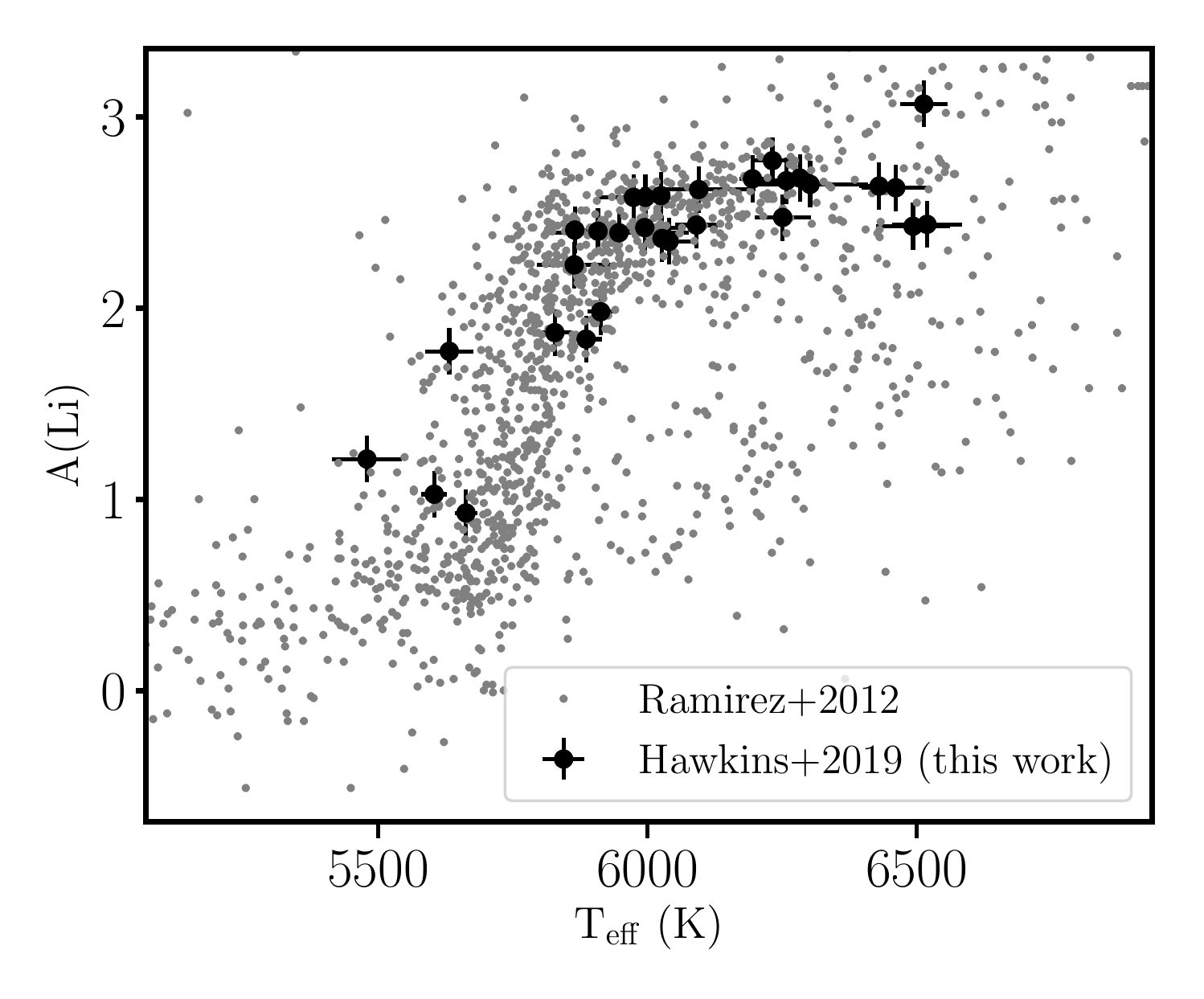}
	\caption{The abundance of Li, denoted as A(Li), as a function of \teff\ in K for our wide binary stars (shown in black), compared to a sample of stars from the Galactic disk from \protect\cite{Ramirez2012} in in gray. } 
 
	\label{fig:LiTeff}
\end{figure}

We determined the abundance of the light element Li using the absorption feature at 6707.8~\AA. Reassuringly, we find that the abundance of lithium, A(Li), of our stars follows a similar trend with \teff\ as expected for typical FGK dwarf stars, namely that A(Li) tends to decrease with decreasing \teff\ and plateaus above \teff\ $\leq$ 6200~K  \citep[e.g.][]{Ramirez2012}. In Fig.~\ref{fig:LiTeff}, we show the abundance of Li, i.e. A(Li) = [Li/H] + 1.05 \citep[where 1.05 is the solar Li abundance,][]{Asplund2005}, for our wide binary stars in black compared to a sample of stars from the Galactic disk from \cite{Ramirez2012}. This figure illustrates two important points: (1) Our sample of wide binaries follows the typical trend in \teff\ dependent depletion as found in the Galactic disk, and (2) if one selects wide binaries with very different \teff\ it may not be expected for their A(Li), and therefore their [Li/H] or [Li/Fe], to be equal. This also may explain why the dispersion in $\Delta$[Li/H] (and subsequently $\Delta$[Li/Fe]) abundance ratios are larger than for other elements. 

\begin{figure}
	 \includegraphics[width=1\columnwidth]{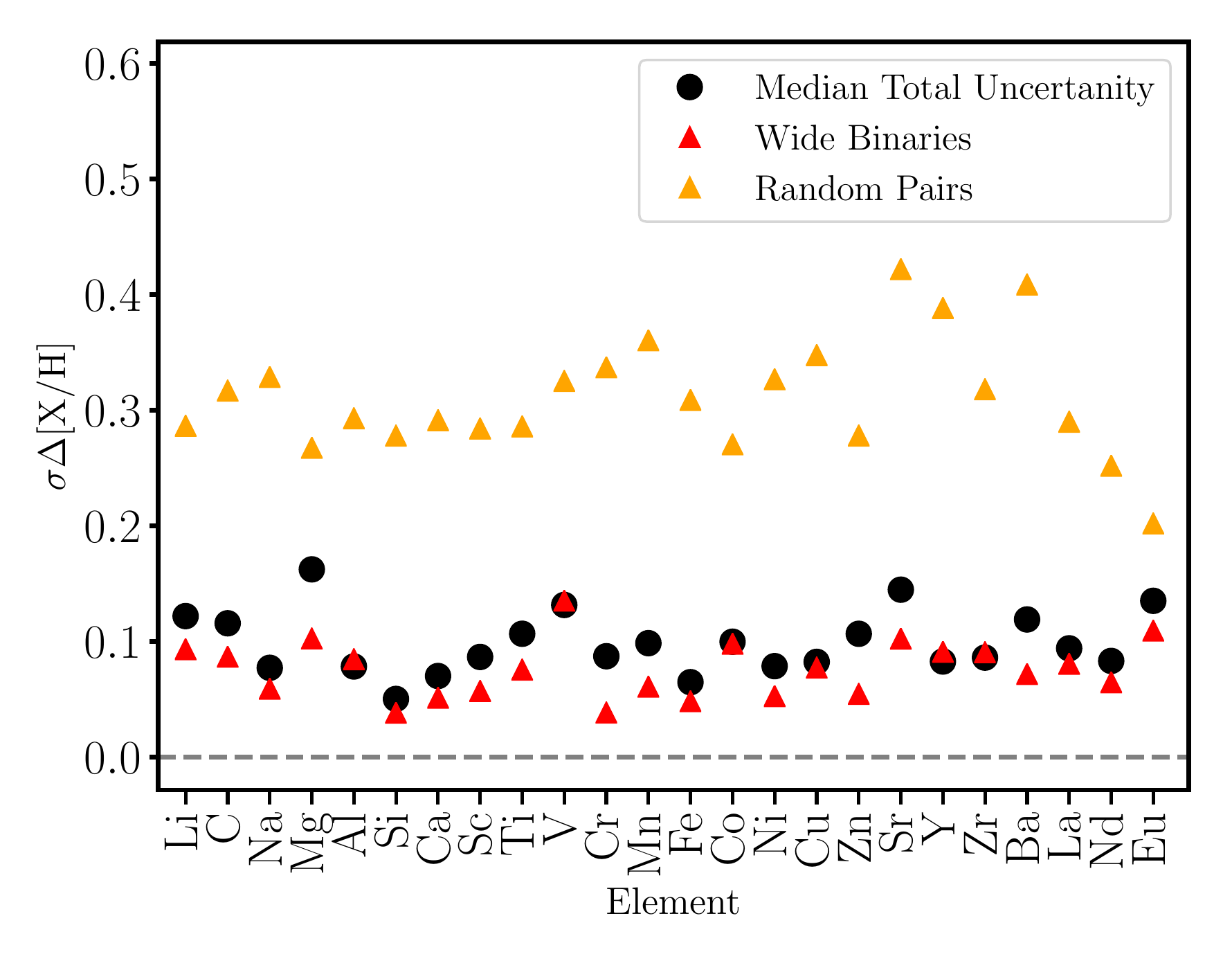}
	\caption{The dispersion in the difference of [X/H] abundance ratios between wide binaries (red triangles) compared to random pairs of stars (orange triangles). For reference, the typical uncertainty in each element is shown as black circles.} 
 
	\label{fig:dispersions}
\end{figure}

\begin{figure}
	 \includegraphics[width=1\columnwidth]{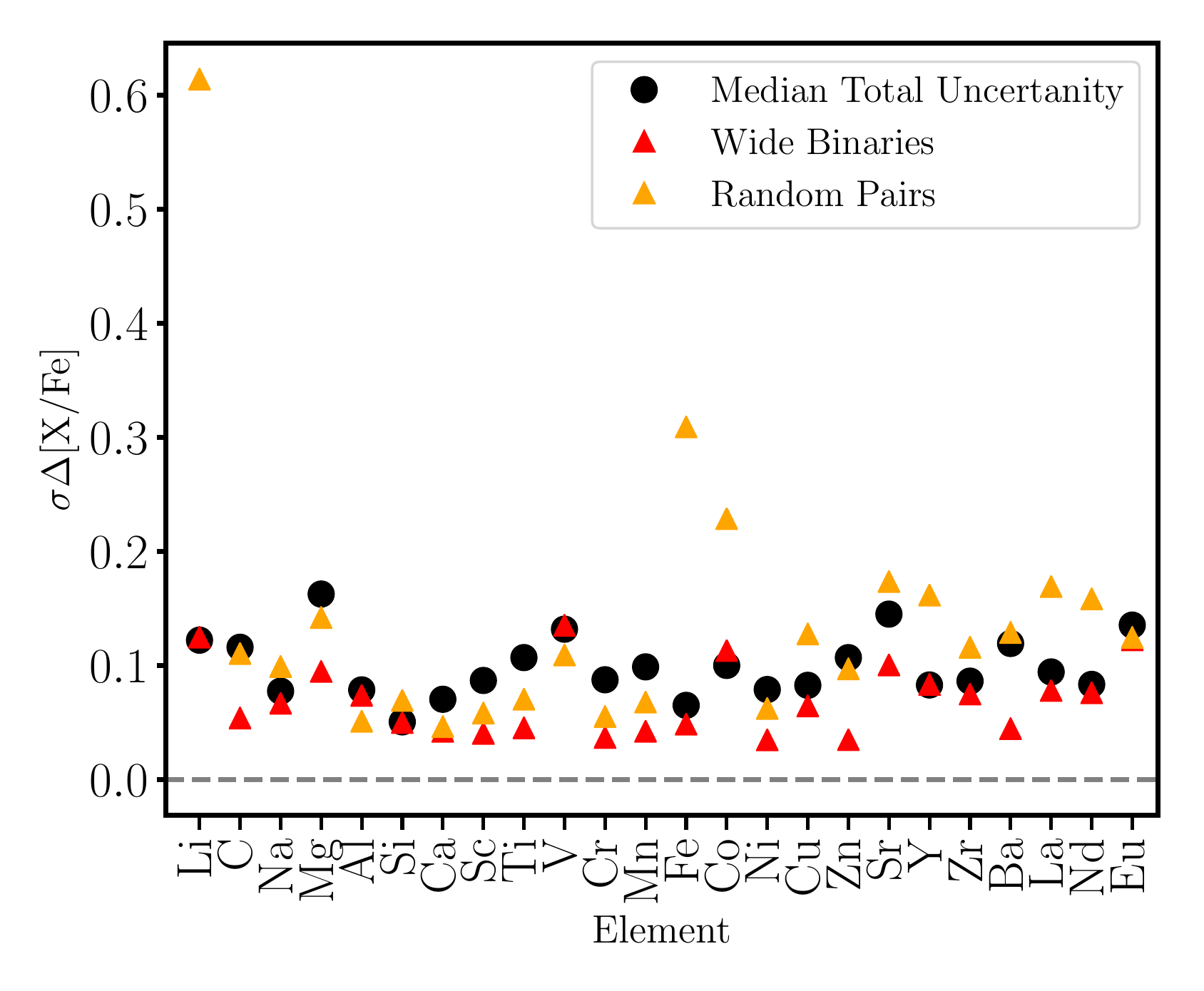}
	\caption{The same as Fig.~\ref{fig:dispersions} except for the dispersion in the difference of [X/Fe] abundance ratios instead. As above, for Fe we show the [Fe/H] for comparison. } 
 
	\label{fig:dispersions_fe}
\end{figure}

Furthermore, we find that the typical difference in Li between the two wide binary pairs is $\Delta$A(Li) = 0.00 with a dispersion of 0.09~dex.  For the purposes of this discussion, we show in Fig.~\ref{fig:dispersions} the dispersion in the difference of [X/H], i.e. $\sigma\Delta$[X/H], as red triangles. Also shown in Fig.~\ref{fig:dispersions} is the typical total uncertainty (as black circles) and the dispersion in the difference of [X/H] for random pairs (orange triangles) instead of wide binaries (red triangle). In Fig.~\ref{fig:dispersions_fe}, we also show the dispersion in the difference of [X/Fe], i.e. $\sigma\Delta$[X/Fe], for the wide binaries in this work (red triangles), the random pairs of stars (as orange triangles). Similar to Fig.~\ref{fig:dispersions}, we also show the typical uncertainty in [X/Fe], which we approximate as the uncertainty in [X/H] added in quadrature with the uncertainty in [Fe/H] for each star. We note here that this analysis does not account for possible covariances between the uncertainties in [X/Fe] and the stellar parameters. Therefore we caution that the uncertainties quoted in Fig.~\ref{fig:dispersions_fe} are conservative.  While we add this figure for completeness, we additionally caution that two random stars not born together as a pair in the Galactic thin disk can have very different \feh\ but very similar [X/Fe], because the dynamic range in [X/Fe] is on the same order (e.g. $\sim$0.10 dex)  as the uncertainty in [X/Fe] (also on the order of $\sim$0.10~dex). This is why [X/H] is critically important for the purposes of chemical tagging. 

These two figures together indicate that the dispersion in the difference of [Li/Fe] is slightly larger compared to the dispersion in  $\Delta$[Li/H] for the wide binaries, which can be explained by increased uncertainties in [Li/Fe]. However for the random pairs, the dispersion in $\Delta$[Li/Fe] is significantly larger than for $\Delta$[Li/H]. This can be attributed both to (1) increased uncertainties in [Li/Fe] compared to [Li/H] and (2) the fact that the Li abundance depends systematically on \teff\ (e.g. Fig.~\ref{fig:LiTeff}) and the random pairs have a large dispersion in \teff\ compared to the wide binaries. This \teff\ dependent Li depletion illustrates why Li should not generally be used for chemical tagging. 

Interestingly, the typical total internal uncertainty in Li is approximately 0.12~dex. We remind the reader this value is derived by adding in quadrature the standard error in the mean of the line-by-line abundances and the sensitivities of the abundance to the stellar parameters. This is in contrast to $\Delta$A(Li) = --0.09 with a significantly larger dispersion of 0.29~dex if one compares each star with the closest star on the CMD, which is not its binary companion. We remind the reader this comparison is a way to quantify the expected difference between random field stars of similar \teff\ and \logg\ internally using the results from our spectra. 

C is a light element and is determined using a combination of molecular features (namely CH) and  two atomic features \cite[][]{Nissen2014}. Using these C features, we were able to derive abundances for [C/H] and find that the typical difference between the two stars in the wide binary pairs is $\Delta$[C/H] ($\Delta$[C/Fe]) = 0.02 (0.02)\footnote{For the purposes of this discussion, and unlike many studies, we will note the difference in [X/Fe] in the parentheses along with the differences in [X/H].} with a dispersion of 0.09 (0.05)~dex in [X/H] ([X/Fe]). This is in contrast to $\Delta$[C/H] = --0.03 (0.02) with a larger dispersion of 0.32 (0.11)~dex if we were to compare each star with the closest star on the CMD, which is not its binary companion. The median total uncertainty in [C/H] is 0.12~dex, dominated by the propagated uncertainty in [C/H] with respect to \teff. While there is a noticeable and significant spread of 0.32~dex in $\Delta$[C/H] when comparing random stars of similar spectral type, we find an offset and spread well below the uncertainty for wide binary pairs consistent with wide binaries originating most often in clouds homogeneous in C. 

Na, Al, Sc, Cu, and V are all odd-Z elements. We determined the abundance of Na in each star using up to 4 absorption features. We find that the median difference in [Na/H, Fe] is 0.00 (0.00)~dex with a dispersion of 0.06 (0.07)~dex. This is in contrast to significantly larger differences, $\Delta$[Na/H] = 0.06 (-0.01) with a dispersion of 0.33 (0.10)~dex for the random pairs. For reference, the median total uncertainty in [Na/H] is 0.07~dex. 


Similarly, for Al, we  use up to 2 absorption features. The typical uncertainty in [Al/H] is $\sim$0.08~dex. For the wide binary systems studied here we find that the median $\Delta$[Al/H] = 0.01 (0.00)~dex and a dispersion in the difference in [Al/H], i.e. $\sigma\Delta$[Al/H], of 0.08 (0.08)~dex. Similar results are also found in Sc where the $\Delta$[Sc/H] = 0.03 (0.01) with a dispersion of 0.06 (0.04)~dex. Note that the typical uncertainty in [Sc/H] is 0.09~dex. Significantly larger dispersion in $\Delta$[Al, Sc/H] are detected for random pairs of stars.

Cu, much like the other odd-Z elements, we find to have a median difference in $\Delta$[Cu/H, Fe] is 0.00 (0.00)~dex with a dispersion in the difference of  $\sigma\Delta$[Cu/H, Fe] is 0.08 (0.06)~dex. With a median total uncertainty in [Cu/H] of 0.08~dex, the differences we find between wide binary pairs in [Cu/H, Fe] is likely a result of measurement uncertainty. On the other hand, for the random pairs of stars the median difference in $\Delta$[Cu/H, Fe] is 0.00 (0.00)~dex with a dispersion in the difference of  $\sigma\Delta$[Cu/H, Fe] is 0.35 (0.13)~dex. The latter indicating that there are measurable differences in [Cu/H, Fe] between random pairs of stars unlike for the wide binaries. 

We determined the abundance of [V/H] using up to 9 absorption lines, which tend to have a relatively large scatter. This is evident by the 0.13~dex median uncertainty in [V/H] across all stars. Despite this, we find that the typical difference in [V/H] for the wide binaries studied here is 0.03 (0.00)~dex with a dispersion of 0.14 (0.14)~dex. As with the other elements, the median difference in [V/H] for the `random pairs'  is 0.06 (0.02)~dex with a significant dispersion of 0.33 (0.11)~dex.  

In each of the light and odd-Z elements, it can be summarized that the median difference in the abundance of component A relative to component B of the wide binary is consistent with zero and has a dispersion less than the typical uncertainty. This is not the case for when we compare each star to its closest {\it non-companion} star in colour-magnitude space.  This result is consistent with other studies \citep[e.g.][]{Andrews2017, Andrews2019}, which find that typical variations in the odd-Z elements are consistent with measurement uncertainties. Of all the odd-Z elements, V  is the species where we observe the largest variation for both the wide binaries and the random pairs. This result is likely due to larger uncertainty with which V is measured.

\subsection{$\alpha$ elements (Mg, Si, Ca)} \label{sub:alpha}

The $\alpha$ elements are those formed during the successive fusion of helium nuclei during the later stages of quasistatic nuclear burning in the inner regions of evolved massive stars. Additionally, Mg and Si play a major role in forming rocks for planets. These elements, which include Mg, Si, and Ca are thought to be dispersed into the interstellar medium by Type~II supernovae. If wide binary systems are formed from chemically homogeneous, well-mixed, turbulent gas then it is expected that the differences in [Mg, Si, Ca/H] between the two stars in the binary system should be consistent with either the measurement uncertainty or the abundance spread for randomly chosen pairs. In Fig.~\ref{fig:XHFediff} we show the difference in  [Mg, Si, Ca/H]  (in the top panel) and  [Mg, Si, Ca/Fe] (in the bottom panel) for the \numpair\ binary pairs observed in this work (orange) and the differences in these abundance ratios for each star and the closest star in the color-magnitude diagram that is not its companion (cyan).  

We find the median difference in [Mg/H] to be on the order of 0.04~(0.02)~dex with a dispersion equal to 0.10 (0.09)~dex. We note however that the typical uncertainty in [Mg/H] is on the order of $\sim$0.15~dex. This is driven by the difficulty in the measurement of Mg which tends to be based on relatively strong lines in these spectral types. Despite this, the differences in [Mg/H, Fe] for the wide binaries are consistent with arising from measurement uncertainty. On the other hand, when comparing random parings of stars we find $\Delta$[Mg/H] = 0.02 with a dispersion that is nearly 3 times larger ($\sigma\Delta$[Mg/H] = 0.29~dex).  A larger dispersion is also observed for [Mg/Fe], where the  dispersion in $\Delta$[Mg/Fe] is 0.09~dex for wide binaries compared to 0.15~dex for random pairs.

The remaining $\alpha$ elements (i.e. Si, Ca) all have median differences in both [X/H] and [X/Fe] less than 0.02~dex. As shown in Fig.~\ref{fig:dispersions} and Fig.~\ref{fig:dispersions_fe}, the dispersion in the difference in [X/H] for the wide binaries are 0.04 and 0.05 for Si, Ca, respectively. These values are similarly 0.04 dex for the dispersion of $\Delta$[Si, Ca/Fe].  For reference the typical total uncertainties in the elements are 0.04, 0.07~dex for [Si/H] and [Ca/H], respectively. While the median difference in [X/H] are slightly larger for the random pairs of stars ($\Delta$[X/H] $\sim$ 0.05~dex) compared to the wide binaries, the dispersion of  the difference in the [X/H] is several times larger ($\sigma\Delta$[Ca, Si/H] $\sim$ 0.30~dex). Such a large dispersion cannot be accounted for by measurement uncertainties alone in the case of the random pairs. Additionally,  Fig.~\ref{fig:dispersions_fe}  illustrates that for the $\alpha$-elements the dispersions in $\Delta$[Mg, Si, Ca/Fe] are generally larger for the random pairs compared to the wide binaries.

Across all of the $\alpha$ elements, we find that for the wide binary systems the differences in [Mg, Si, Ca/H] abundance ratios are explained by the measurement uncertainties. This is also the case for [Mg, Si, Ca/Fe]. Among all of the $\alpha$ elements, we find the largest differences in the Mg, with $\Delta$[Mg/H] = 0.05 $\pm$ 0.10~dex. This is likely  driven by the larger uncertainties. Interestingly, this echos recent results from \cite{Andrews2019}, who use the fourteenth data release from the infrared APOGEE survey \citep{Holtzman2018} to study the chemical homogeneity of 31~wide binary systems. They concluded that of all of the $\alpha$ elements only Mg potentially shows genuine abundance differences, though they note that this could be a result of the uncalibrated \logg\ for APOGEE dwarf stars. For the remaining elements, in line with recent work \citep[e.g.][]{Andrews2017, Andrews2019}, we find that the wide binary systems are consistent at a level below $\sim$0.05~dex.

\subsection{Fe-peak elements (Ti, Cr, Mn, Co, Ni, Zn)} \label{sub:Fepeak}

Other than iron, Ti, Cr, Mn, Co, Ni, and Zn represent elements near the Fe-peak. We note that Ti is often is classified in the literature both as an $\alpha$ and Fe-peak element. Ti is not directly formed through the successive addition of $\alpha$ particles, but rather through as a decay product of $^{48}$Cr \citep[e.g.][]{Curtis2019}. Its observed chemical evolution displays similarity to both $\alpha$ and Fe-peak elements, but here we classify it as an Fe-peak element. These elements are thought to be formed and dispersed into the interstellar medium primarily through Type~Ia supernova explosions \citep[e.g.][and references therein]{Iwamoto1999, Kobayashi2006,Kobayashi2011, Nomoto2013}. We find that in all Fe-peak elements the median differences in abundance ratios between both components of a wide binary, $\Delta$[Ti, Cr, Mn, Co, Ni, Zn/H] and $\Delta$[Ti, Cr, Mn, Co, Ni, Zn/Fe], are less than 0.03~dex. The dispersion in the $\Delta$[X/H] for the wide binaries are  0.08, 0.04, 0.06, 0.09, 0.05, 0.05~dex for Ti, Cr, Mn, Co, Ni, and Zn, respectively. These are compared to the typical total uncertainties in these elements, which are 0.10, 0.09, 0.10, 0.10, 0.08, and 0.11~dex  for Ti, Cr, Mn, Co, Ni, and Zn, respectively. 

These values are sufficiently different and smaller than for random pairs of stars which are similar in spectral type. For example, while we find the median offset in  $\Delta$[Ti, Cr, Mn, Co, Ni, Zn/H] is similar to wide binaries the dispersions are significantly larger ($\sigma\Delta$[X/H]$>$0.30~dex). This is also the case for $\sigma\Delta$[Ti, Cr, Mn, Co, Ni, Zn/Fe], where they are typically 2-3 times larger for random pairs compared to wide binary systems. 

These results suggest that in each of the Fe-peak elements the wide binary systems are chemically homogeneous, having differences below 0.03~dex. Additionally, the dispersions in the differences of the abundance ratios, $\sigma\Delta$[X/H, Fe], are consistent with the uncertainty for each element indicating that the distribution in abundance differences that we observe in our sample of wide binary stars is likely due to measurement uncertainty. We note that this is not the case if we were to compare `random pairs' of stars in similar (and not similar) parts of the color magnitude diagram. 

This result is consistent with earlier works \citep[e.g.][]{Desidera2004, Desidera2006}, which suggest wide binary systems tend to be chemically homogeneous in Fe and Fe-peak elements. This is contrary to the results of a limited sample of 11 wide binaries studied using the GALAH survey \citep{Simpson2018}. However, for a handful of stars in our sample we do find potentially significant ($\Delta$[Fe/H] $\sim$0.10--0.15~dex) differences in \feh. These systems tend to also be  enhanced in the other Fe-peak elements. While this is not common, this can be indicative of the existence or accretion of planetary material in these systems. A difference of $\Delta$\feh\ $\leq$ 0.14~dex in Fe-peak elements has been found in other systems \citep[e.g. Koronos, HAT-P-4, HIP 68468, and others,][]{Oh2017, Saffe2017, Melendez2017}. Furthermore, \cite{Simpson2018} used the GALAH survey and found a higher prevalence ($\sim$60\%) of systems where the binary pairs differed in \feh\ by $\sim$0.10~dex or more.

\subsection{Neutron capture elements (Sr, Y, Zr, Ba, La, Nd, Eu)} \label{sub:ncap}
The neutron capture elements include those that are formed though slow (s-process) or rapid (r-process) successive neutron capture. These heavy elements are produced and dispersed into the interstellar medium in a variety of ways \citep[e.g. asymptotic giant branch stars, supernova, neutron star - neutron stars mergers etc.][]{Nomoto2013}. We measure the elemental abundances of both s-process (Sr, Y, Zr, Ba, La, and Nd) and r-process (Eu) elements.

We find that the median difference in $\Delta$[Sr, Y,  Zr, Ba, La, Nd, Eu/H] and their corresponding abundance ratios with Fe, are always less than 0.03~dex. The dispersion in the difference of the [X/H] ratios for the neutron capture elements, i.e. $\sigma\Delta$[Sr, Y, Zr, Ba, La, Nd, Eu/H] , are found to be 0.10, 0.09, 0.10, 0.11, 0.08, 0.07, 0.11, respectively.  The typical uncertainties in these elements are 0.08~dex for Y and Nd, 0.09 for Zr, Ba, and La, and 0.13 and 0.15~dex for Eu and Sr, respectively. These typical uncertainties are slightly larger than for the $\alpha$ and Fe-peak elements due to the difficulty of measuring these elements. This is likely a result of the lack of many quality absorption features for several neutron capture elements. Despite this, we find that the distribution in $\Delta$[Sr, Y, Zr, Ba, La, Nd, Eu/H] are very close to what is expected due to measurement uncertainties. 

Similar to the other elemental families, we find that random pairs of (non-companion) stars, whether chosen in a completely random way or selected to be in a similar part of the color-magnitude plane, are chemically different. While the median $\Delta$[Sr, Y, Zr, Ba, La, Nd, Eu/H] is lower than 0.08~dex for each element, the dispersion in the difference is significantly larger ($\sigma\Delta$[Sr, Y, Zr, Ba, La, Nd, Eu/H] $\sim$0.35--0.40~dex). The dispersion in the difference of the [X/Fe] ratios for the neutron capture elements, i.e. $\sigma\Delta$[Sr, Y, Zr, Ba, La, Nd, Eu/Fe], are found to range between 0.08--0.12~dex with typical value of 0.10~dex for the case of the wide binaries. For the random pairs,  the dispersion in the difference of the [X/Fe] ratios for the neutron capture elements, i.e. $\sigma\Delta$[Sr, Y, Zr, Ba, La, Nd, Eu/Fe], are found to range between 0.012-0.17~dex with typical value of 0.15~dex. As before, the random pairs of stars tend to have larger dispersions in $\Delta$[Sr, Y, Zr, Ba, La, Nd, Eu/Fe] compared to the wide binaries.
\\ \\
The chemical differences of neutron capture elements has been studied in solar twins and wide binary systems \citep[e.g.][and others]{Melendez2009, Teske2016, Melendez2017,Simpson2018}. These studies and surveys have shown that generally the neutron capture elements can vary as much as 0.1~dex between the two stars in wide binary systems, though often vary at much lower levels. Theoretical work on the homogeneity of the gas clouds from which wide binaries could form seem to suggest that if there are differences in the chemical abundance ratios, they should be as large ($\Delta$[X/H] $\leq$0.20~dex) in the neutron capture elements \citep[e.g.][]{Krumholz2018} as in all other elements depending on their formation channel. Our results indicate that while there are larger differences in the neutron capture elements (especially Eu), this is mostly likely due to the larger uncertainties with which we can measure these elements.

\subsection{Chemical Inhomogeneity and the Prospects for Chemical Tagging}
In the above sections, we present the chemical abundance distributions for light/odd-Z (section~\ref{sub:light}), $\alpha$ (section~\ref{sub:alpha}), Fe-peak (section~\ref{sub:Fepeak})  and neutron capture (section~\ref{sub:ncap}) elements in the \numpair\ wide binaries. We also place the key abundance differences in the context of other studies. In order for chemical tagging to be viable, one would expect that wide binary systems that formed together are chemically homogeneous within the precision of measurement for each elemental abundance ratio. This is however not expected for random pairs of field stars, whether selected to be similar in \teff\ and \logg\ or not. It is also not expected that wide binary systems formed through dynamical effects (resonance structure or tidal capture) be chemically alike. 

We find that most (20/25 systems) of the wide binary stellar systems studied here are of equal metallicity, within the typical uncertainties, with $\Delta$\feh = $0.01 \pm$ 0.02~dex. This result echos previous studies which have showed that wide binaries \citep{Gizis1997, Gratton2001, Martin2002, Desidera2004, Desidera2006} or the larger open cluster cousins \citep[e.g.][]{Bovy2016, Liu2016, Ness2018} are chemically consistent to a level of 0.02--0.04~dex, but that small variations could be present at below these levels. 

Of the 25 systems studied, 5 have $\Delta$\feh $>$ 0.10~dex. These systems also tend to have the largest differences in the remaining elements studied. The reason for these observed chemical abundance differences can be related to several effects including the ingestion of (rocky) planetary material \citep[e.g.][]{Melendez2009, Oh2017}, atomic diffusion \citep[e.g.][]{Dotter2017}, mass transfer from the companion \citep[e.g.][]{Hansen2015}, or the formation of wide binary systems through exchange scattering, among other things. It is also possible that some of the chemically-discrepant pairs in our sample only appear as such because the wide binary is really a hierarchical triple, with one resolved component having an unresolved companion which contributes to the spectrum. Such systems are reasonably common -- \citet{ElBadry2018a} estimated that roughly 20\% of the wide binaries in their catalog contain a component with an unresolved companion bright enough to contribute substantially to the spectrum -- and such unresolved companions can bias the derived abundances at the 0.1 dex level \citep{ElBadry2018c}. We do not attempt to determine which of these may be the cause of the metallicity discrepancy for the pairs where $\Delta$\feh\ $>$ 0.10~dex. However, we note that we did explore the detailed differences in $\Delta$[X/H] with respect to the condensation temperature \citep[T$_c$,][]{Lodders2003}. Correlations between T$_c$ and the enhancement of [X/H] is thought to be indicative of rocky planetary accretion \citep[e.g.][ and references therein]{Melendez2009, Oh2017}. In some cases, we see a reasonable trend between indicative of the accretion of rocky material, but not in all cases. This warrants a separate study. We note that the likelihood of forming these systems through exchange scattering is low enough to be negligible in the field population \citep[for example, see equation 8 of][]{Oh2017}. It is also not likely to be a result of atomic diffusion since these stars are close in \teff\ and \logg. It is clear, however, that the bulk of the wide binaries ($\sim$80\%) are in fact chemically homogeneous. It will be critical to observe more pairs, either through dedicated observing campaigns or through large spectroscopic surveys, to (i) identify the fraction of chemically dissimilar wide binaries and (ii) to define the parameter space where chemically dissimilar wide binaries  are more likely to be found.




\section{Summary} \label{sec:summary}
Wide binary systems represent a unique testing ground for not only the concept of chemical tagging but also for the methods used in the characterization of difficult-to-analyze stars (such as M-dwarfs or white dwarfs). One of the primary underlying assumptions of these techniques is that stars born together are chemically homogeneous. For chemical tagging to work,  one would expect  no differences in the observed [X/H] or [X/Fe] abundance ratios measured in both components of a wide binary systems born from the same gas cloud.

Early work done on wide binaries suggested that they may in fact be chemically homogeneous in \feh, but other elements were still in question \citep[e.g.][]{Martin2002, Dotter2003, Desidera2004, Desidera2006}. Recently, the advent of the large astrometric surveys, particularly the \gaia\ mission, have enabled the discovery of many new wide binary systems \citep[e.g.][]{Andrews2017, Oh2017, ElBadry2018a} with which we can further test the prediction. \cite{Oh2017}, made it clear that not all wide binary systems are chemically homogeneous and can be dissimilar by as much as 0.10~dex. Follow-up work by \cite{Simpson2018} indicated that this may be as prevalent as $\sim$60\% of wide binaries.

In this work, we obtained high-resolution (R$\sim$60,000) high signal-to-noise ratio (SNR $\geq$ 60 pixel$^{-1}$) spectra of  \numstar\ stars making up \numpair\ wide binary pairs (section~\ref{subsec:spec}). These wide binaries were identified using \gaia\ DR2 and selected from the catalogue presented in \cite{ElBadry2018a}. Using the collected spectra, we derived the stellar atmospheric parameters (\teff, \logg, \feh, \vmicro) for each stars and chemical abundances for \numel\ species across the light/odd-Z (Li, C, Na, Al, Sc, V, Cu, $\alpha$ (Mg, Si, Ca), Fe-peak (Ti, Cr, Mn, Fe, Co, Ni, Zn), and neutron capture (Sr, Y, Zr, Ba, La, Nd, Eu) families using the BACCHUS stellar parameter and abundance pipeline (section~\ref{sec:SP}).

We compared both the [X/H] and [X/Fe] abundance ratios of both stars in the wide binary pair (their difference can be found in Fig.~\ref{fig:XHFediff} and discussed in more detail in section~\ref{sec:results}). Results indicate that 80\% of the sample (i.e. 20/25 wide binaries studied here) have been found to have equal \feh\ to within $\sim$0.02~dex while the remaining five systems have $\Delta$\feh $\sim$0.10~dex. In most of the elements studied the distribution of the difference in abundance ratios (in both $\Delta$[X/H] and $\Delta$[X/Fe]) between wide binary pairs are consistent with  measurement uncertainty (which for most elements is on the order of $\sigma$[X/H] $\leq$ 0.08~dex across all elements). We also compared these to the differences in chemical abundance ratios between each star and the closest stars on the color-magnitude diagram which is not its binary companion, as well as random pairings of these field stars. As expected, wide binary systems are far more homogeneous compared to simple random pairings of field stars. 

These results enable us to conclude that wide binary systems are likely to be chemically homogeneous though in some cases they may not be, consistent with other works \citep[e.g.][]{Desidera2006, Oh2017, Andrews2017, Andrews2019}. This is encouraging for chemical tagging at the level of $\sim$0.08~dex for most elements. We predict that chemically inhomogeneous wide binaries may occur on the order $\sim$20\% of the time. Larger samples of wide binaries, either through large spectroscopic surveys or better even high-resolution followup, will enable us to test this prediction \citep[e.g.][]{Andrews2017, Andrews2019}. These samples will enable not only an extension of the current work, but may also enable us to determine under which conditions binary stellar systems are least likely to be chemically homogeneous, which will be critical to the success of chemical tagging. 

\section*{Acknowledgements}
{\small 
We thank the referee whose detailed and thorough report improved this manuscript significantly. This paper includes data taken at The McDonald Observatory of The University of Texas at Austin. We thank the staff at McDonald Observatory  for making this project possible. KH has been partially supported by a TDA/Scialog grant funded by the Research Corporation and a Scialog grant funded by the Heising-Simons Foundation. KH acknowledges support from the National Science Foundation grant AST-1907417. This work was performed, in part, at the Aspen Center for Physics, which is supported by National Science Foundation grant PHY-1607611. D.K and M.T have been supported by Cox Endowment funds through the Board of Visitors of the University of Texas at Austin Department of Astronomy. Y.S.T. is grateful to be supported by the NASA Hubble Fellowship grant HST-HF2-51425.001 awarded by the Space Telescope Science Institute. A.P.J. is supported by NASA through Hubble Fellowship grant HST-HF2-51393.001 awarded by the Space Telescope Science Institute, which is operated by the Association of Universities for Research in Astronomy, Inc., for NASA, under contract NAS5-26555. Support for this work was provided by NASA through Hubble Fellowship grant HST-HF2-51399.001 awarded to J.K.T. by the Space Telescope Science Institute, which is operated by the Association of Universities for Research in Astronomy, Inc., for NASA, under contract NAS5-26555. A.C. thanks the LSSTC Data Science Fellowship Program, which is funded by LSSTC, NSF Cybertraining Grant 1829740, the Brinson Foundation, and the Moore Foundation; Her participation in the program has benefited this work.

This work has made use of data from the European Space Agency (ESA)
mission {\it Gaia} (\url{https://www.cosmos.esa.int/gaia}), processed by
the {\it Gaia} Data Processing and Analysis Consortium (DPAC,
\url{https://www.cosmos.esa.int/web/gaia/dpac/consortium}). Funding
for the DPAC has been provided by national institutions, in particular
the institutions participating in the {\it Gaia} Multilateral Agreement.}

\appendix
\section{Online Tables} 
In order to ensure that this work is not only reproducible but also useful to the community, we provide here two online tables. In Table~\ref{tab:linebyline} we provide a small portion of a much lager table which collects the atmospheric abundances derived in this work for each star, elements, and absorption feature. We choose to only provide a short example of this in  Table~\ref{tab:linebyline} for brevity. We note that the abundances of species $X$, denoted as $\log{A_X}$, are in the usual form where $\log{A_X} = \log{\frac{N_X}{N_H}}$ where $\log{N_H}$ is normalized to 12.00. 

In addition to the abundance information for each star, element and absorption line, we also provide the some basic atomic data (including the wavelength, in \AA, the $\log{gf}$, and the excitation potential, in eV) for each line. References are also provided in the reference key column of Table~\ref{tab:linebyline}, which can be matched to Table~\ref{tab:refkey} for the full citation.  

\begin{table*}
\caption{Line-by-line abundance information} 
\begin{tabular}{ccccccc}
\hline
Name & Element & $\lambda$ & $\log{gf}$ & Reference Key & $\chi$ & $\log{A_X}$ \\ \hline\hline
& & (\AA) &(dex) & & (eV) & (dex)\\
 \hline
 ...&...&...&...&...&...&...\\
WB24B & Mg I & 5711.08 & -1.724 & 1990JQSRT..43..207C & 4.346 & 7.579 \\
WB24B & Mg I & 6318.71 & -2.103 & 1993JPhB...26.4409B & 5.108 & 7.568 \\
WB25A & Mg I & 4730.02 & -2.347 & NIST10 & 4.346 & 7.51 \\
WB25A & Mg I & 5711.08 & -1.724 & 1990JQSRT..43..207C & 4.346 & 7.388 \\
WB25A & Mg I & 6318.71 & -2.103 & 1993JPhB...26.4409B & 5.108 & 7.324 \\
WB25B & Mg I & 4730.02 & -2.347 & NIST10 & 4.346 & 7.53 \\
WB25B & Mg I & 5711.08 & -1.724 & 1990JQSRT..43..207C & 4.346 & 7.381 \\
WB01A & Al I & 5557.06 & -2.104 & 1995JPhB...28.3485M & 3.143 & 6.794 \\
WB01A & Al I & 6696.02 & -1.569 & GESG12 & 3.143 & 6.837 \\
WB01B & Al I & 5557.06 & -2.104 & 1995JPhB...28.3485M & 3.143 & 6.763 \\
WB01B & Al I & 6696.02 & -1.569 & GESG12 & 3.143 & 6.829 \\
WB02A & Al I & 6696.02 & -1.569 & GESG12 & 3.143 & 6.18 \\
WB03B & Al I & 6696.02 & -1.569 & GESG12 & 3.143 & 5.952 \\
WB04A & Al I & 5557.06 & -2.104 & 1995JPhB...28.3485M & 3.143 & 6.309 \\
WB04A & Al I & 6696.02 & -1.569 & GESG12 & 3.143 & 6.493 \\
WB04B & Al I & 5557.06 & -2.104 & 1995JPhB...28.3485M & 3.143 & 6.432 \\
WB04B & Al I & 6696.02 & -1.569 & GESG12 & 3.143 & 6.476 \\
WB05A & Al I & 5557.06 & -2.104 & 1995JPhB...28.3485M & 3.143 & 6.57 \\
WB05B & Al I & 5557.06 & -2.104 & 1995JPhB...28.3485M & 3.143 & 6.641 \\
WB05B & Al I & 6696.02 & -1.569 & GESG12 & 3.143 & 6.409 \\
...&...&...&...&...&...&...\\
\hline \hline
\label{tab:linebyline}
\end{tabular}
\raggedright

NOTE: This is a cut out of a long table which includes the of the derived stellar abundances ($\log{A_X}$, last column) for each line and elemental species and star discussed in this work. The name of the star is in column~1 while each elemental species, its wavelength, its $\log{gf}$ can be found in columns~2-4, respectively. We also indicate reference (through the reference key) where that atomic data (specifically the $\log{gf}$) was taken sourced. The reference key can be matched to exact reference through Table~\ref{tab:refkey}. 
\end{table*}

\clearpage
\onecolumn
\begin{longtable}{|l|l}

 \caption{Atomic Data References \label{tab:refkey}}\\
 \hline
 \multicolumn{2}{|l|}{}\\
 \hline
 Reference Key & Reference \\
 \hline
 \endfirsthead

 \endhead
 
 \hline
 \endfoot

2007AA...472L..43B & \cite{2007AA...472L..43B} \\
BGHL & \cite{BGHL} \\
BK & \cite{BK} \\
BKK & \cite{BKK} \\
BL & \cite{BL} \\
BWL & \cite{BWL} \\
CC & \cite{CC} \\
DLSSC & \cite{DLSSC} \\
FMW & \cite{FMW} \\
GARZ & \cite{GARZ} \\
GESB82c & \cite{GESB82c} \\
GESB82d & \cite{GESB82d} \\
GESB86 & \cite{GESB86} \\
GESG12 & \cite{GESG12} \\
GESHRL14 & \cite{GESHRL14} \\
GESMCHF & \cite{GESMCHF} \\
HLSC & \cite{HLSC} \\
K03 & \cite{K03} \\
K07 & \cite{K07} \\
K10 & \cite{K10} \\
K12 & \cite{K12} \\
K13 & \cite{K13} \\
KR & \cite{KR} \\
LBS & \cite{LBS} \\
LD & \cite{LD} \\
LGWSC & \cite{LGWSC} \\
LNAJ & \cite{LNAJ} \\
LWHS & \cite{LWHS} \\
LWST & \cite{LWST} \\
MRW & \cite{MRW} \\
MW & \cite{MW} \\
NIST10 & \cite{NIST10} \\
NWL & \cite{NWL} \\
PGBH & \cite{PGBH} \\
PRT & \cite{PRT} \\
PTP & \cite{PTP} \\
RU & \cite{RU} \\
S & \cite{S} \\
SK & \cite{SK} \\
SLS & \cite{SLS} \\
SR & \cite{SR} \\
WBW & \cite{WBW} \\
Wc & \cite{Wc} \\
1970AA.....9...37R & \cite{1970AA.....9...37R} \\
1980AA....84..361B & \cite{1980AA....84..361B} \\
1980ZPhyA.298..249K & \cite{1980ZPhyA.298..249K} \\
1982ApJ...260..395C & \cite{1982ApJ...260..395C} \\
1982MNRAS.199...21B & \cite{1982MNRAS.199...21B} \\
1983MNRAS.204..883B & \cite{1983MNRAS.204..883B} \\
1984MNRAS.207..533B & \cite{1984MNRAS.207..533B} \\
1984MNRAS.208..147B & \cite{1984MNRAS.208..147B} \\
1986MNRAS.220..289B & \cite{1986MNRAS.220..289B} \\
1989AA...208..157G & \cite{1989AA...208..157G} \\
1989ZPhyD..11..287C & \cite{1989ZPhyD..11..287C} \\
1990JQSRT..43..207C & \cite{1990JQSRT..43..207C} \\
1992AA...255..457D & \cite{1992AA...255..457D} \\
1993JPhB...26.4409B & \cite{1993JPhB...26.4409B} \\
1995JPhB...28.3485M & \cite{1995JPhB...28.3485M} \\
1998PhRvA..57.1652Y & \cite{1998PhRvA..57.1652Y} \\
1999ApJS..122..557N & \cite{1999ApJS..122..557N} \\
2003ApJ...584L.107J & \cite{2003ApJ...584L.107J} \\
2009AA...497..611M & \cite{2009AA...497..611M} \\
2013ApJS..205...11L & \cite{2013ApJS..205...11L} \\
2013ApJS..208...27W & \cite{2013ApJS..208...27W} \\
2014ApJS..211...20W & \cite{2014ApJS..211...20W} \\
2014MNRAS.441.3127R & \cite{2014MNRAS.441.3127R} \\
\hline\hline
\end{longtable}
\twocolumn

\bibliography{bibliography}
\bsp	
\label{lastpage}
\end{document}